\documentclass{emulateapj}

\slugcomment{To appear in the Astrophysical Journal Supplement Series}
\bibpunct[, ]{(}{)}{;}{a}{}{,}
\newcommand{\cardelli}{Galactic~extinction~law}
\newcommand{\cardellitab}{Galactic}
\newcommand{\beq}{\begin{equation}}
\newcommand{\eeq}{\end{equation}}

\shorttitle{Extinction curves of lensing galaxies}
\shortauthors{El\'\i asd\'ottir et al.}

\begin{document}

\title{Extinction curves of lensing galaxies out to $z=1$%
\thanks{Based on observations made with ESO Telescopes at the La Silla or Paranal Observatories under programme IDs 065.O-0666 and 066.A-0264.}
}

\author{\'A.~El\'\i asd\'ottir\altaffilmark{2}, 
J.~Hjorth\altaffilmark{2}, S.~Toft\altaffilmark{3,4}, I. Burud\altaffilmark{5}, D. Paraficz\altaffilmark{2,6}}

\altaffiltext{2}{Dark Cosmology Centre, Niels Bohr Institute, University of
Copenhagen, Juliane Maries Vej 30, DK-2100 Copenhagen \O, Denmark; 
ardis@dark-cosmology.dk.}
\altaffiltext{3}{European Southern Observatory, Karl-Schwarzschild-Stra\ss e 2, 85748 Garching bei M\"{u}nchen, Germany}
\altaffiltext{4}{Department of Astronomy, Yale University, P.O. Box 208101, New Haven, CT 06520-8101}
\altaffiltext{5}{Norwegian Meteorological Institute, P.O. Box 43, Blindern, N-0313 Oslo, Norway}
\altaffiltext{6}{Nordic Optical Telescope (NOT), Apartado 474, 38700 Santa Cruz de La Palma, Canary Islands, Spain}


\begin{abstract}
We present a survey of the extinction properties of ten lensing galaxies, in the redshift range $z=0.04$--$1.01$, using multiply lensed quasars imaged with the ESO VLT in the optical and near infrared.  The multiple images act as `standard light sources' shining through different parts of the lensing galaxy, allowing for extinction studies by comparison of pairs of images.   We explore the effects of systematics in the extinction curve analysis, including extinction along both lines of sight and microlensing, using theoretical analysis and simulations.  In the sample, we see variation in both the amount and type of extinction.  Of the ten systems, seven are consistent with extinction along at least one line of sight.  The mean differential extinction for the most extinguished image pair for each lens is $\bar{A}(V) = 0.56\pm0.04$, using \cardelli~parametrization.  The corresponding mean $\bar{R}_V=2.8\pm0.4$ is consistent with that of the Milky Way at $R_V=3.1$, where $R_V=A(V)/E(B-V)$.  We do not see any strong evidence for evolution of extinction properties with redshift.  Of the ten systems, B1152+199 shows the strongest extinction signal of $A(V)=2.43\pm0.09$ and is consistent with a \cardelli~with $R_V=2.1\pm0.1$.  Given the similar redshift distribution of SN Ia hosts and lensing galaxies, a large space based study of multiply imaged quasars would be a useful complement to future dark energy SN Ia surveys, providing independent constraints on the statistical extinction properties of galaxies up to $z\sim1$.

\end{abstract}




\section{Introduction}

The study of extinction curves of galaxies at high redshift has generated a lot of interest in recent years \citep[see e.g.,][]{riess,  falco1999,goudfrooij,murphy,kann,  goicoechea, york}.  Light reaching us from distant sources is extinguished by dust along its path making it important to correct measurements for the amount and properties of the extinction.  Extragalactic dust extinction can for example affect measurements of Type Ia supernovae (SNe Ia) used to determine various cosmological parameters \citep[e.g.,][]{riess1998, perlmutter} and the star-formation rates for high redshift starburst galaxies which are used as probes of galaxy evolution \citep[see e.g.,][]{madau}.  Yet, even if dust properties and thus extinction may vary with redshift and environment,  an average \cardelli~is often applied when calibrating extragalactic data due to the lack of knowledge of the extinction properties of higher redshift galaxies.   

Traditionally, extinction curves are measured by comparing the spectra of two stars of the same spectral type, which have been reddened by different amounts \citep[see e.g.,][]{massa}.  As it becomes significantly harder to measure spectra of individual stars with distance this method is limited in its application to the Milky Way and the nearest galaxies.  The extinction curves of the Milky Way along different lines of sight have been mapped extensively using this method and have been shown to follow an empirical parametric function which depends only on one parameter, $R_V=A(V)/E(B-V)$, where $A(\lambda)$ is the total extinction at wavelength $\lambda$ and $E(B-V)=A(B)-A(V)$ \citep{cardelli}.  The mean value of $R_V$ in the Milky Way is $3.1$ \citep{cardelli} but for different lines of sight the value ranges from as low as $R_V\approx1.8$ toward the Galactic bulge \citep{udalski} and as high as $R_V\approx5.6$--$5.8$ \citep{cardelli,fitzpatrick}.  A lower $R_V$ corresponds to a steeper rise of the extinction curve into the UV, whereas it has little effect on the extinction in the infrared.  

Extinction curves have also been obtained for the Small and Large Magellanic Clouds (hereafter, SMC and LMC, respectively) and M31 using this method.  The mean extinction curve of the LMC differs from the \cardelli~in that the bump at $2175$~$\AA$ is smaller by a factor of two (as measured by the residual depth of the bump when the continuum has been extracted) and the curve has a steeper rise into the UV for wavelengths shorter than $2200$~$\AA$ \citep{nandy81}.  The extinction curve of the SMC is well fitted by  an $A(\lambda)\propto\lambda^{-1}$ curve which deviates significantly from the \cardelli~and the LMC extinction for $\lambda^{-1}\ge4$~$\mu m^{-1}$ and in particular shows no bump at $2175$ $\AA$  and a steeper rise into the UV \citep{prevot84}.  \citet{bianchi} found that the extinction of M31 follows that of the average \cardelli.  The various extinction properties shown by these galaxies, especially in the UV and shorter wavelengths, further strengthens the need to find a method to study the extinction curves of more distant galaxies.
 
A few methods have been proposed for measuring extinction curves for more distant galaxies.  One method is basically an extension of the traditional method of comparing stars of the same spectral type to comparing the SNe Ia \citep{riess,perlmutter97}.  The extinction is estimated from comparison with unreddened, photometrically similar SNe Ia.  A subset of SNe Ia, with accurately determined extinctions and relative distances, is then used to further determine the relationship between light and color curve shape and luminosity in the full sample.  SN Ia extinction studies usually give lower $R_V$ values than the mean Galactic value of $R_V=3.1$ \citep{riess1996b, krisciunas, wang2006}.

Quasars with damped Ly$\alpha$ systems (DLAs) in the foreground have also been studied by \citet{pei} and were found to be on average redder than those without.   By comparing the optical depths derived from the spectral indices and the ones derived from excess extinction at the location of the \cardelli~bump they found that their sample of five quasars with DLAs is not consistent with the \cardelli, marginally compatible with the LMC extinction and fully compatible with SMC extinction. \citet{murphy} studied a larger sample of the Sloan Digital Sky Survey quasars with damped Ly$\alpha$ systems in the foreground and found no sign of extinction.  They suggest that the difference between their results and that of \citet{pei} may be due to the small number statistics in the study by \citet{pei}.  \citet{ellison} also found that intervening galaxies cause a minimal reddening of background quasars in agreement with the results of \citet{murphy} while \citet{york} found $E(B-V)$ of up to $0.085$ for quasars from the Sloan Digital Sky Survey with Mg II absorption.  In their study \citet{york} found no evidence of the $2175$~$\AA$ bump (at variance with \citet{malhotra}) and found that the extinction curves are similar to SMC extinction.  \citet{ostman} studied the feasibility of measuring extinction curves by using quasars shining through galaxies.  For the two such systems which survived their cuts, they argued that the extinction curves in the foreground spiral galaxies were consistent with Galactic extinction.  They further suggested a possible evolution in the dust properties with redshift, with higher $z$ giving lower $R_V$ by studying values obtained from the literature in addition to their own. 

Extinction curves of high redshift galaxies have also been studied by looking at the spectral energy distribution of gamma-ray bursts (GRBs).  For example, \citet{palli} fitted a \cardelli, an SMC and an LMC extinction law to the afterglow of  GRB 030429. The afterglow, at $z=2.66$, was best fit by an SMC like extinction curve with $A(V)=0.34\pm0.04$. \citet{kann} studied the extinction of a sample of 19 GRB afterglows and fitted them to various dust extinction models.  They found that the SMC extinction law was preferred by a great majority of their Golden Sample (seven out of eight) while one afterglow was best fit by a \cardelli~(the other eleven were equally well fit by SMC, LMC and Galactic extinction).  The mean extinction in the $V$-band was $A(V)=0.21\pm0.04$.

\citet{goudfrooij} reviewed the dust properties of giant elliptical galaxies and found that they are typically characterized by small $R_V$ if they are in the field or in loose groups, but that if they are in dense groups or clusters their $R_V$ values are close to the mean Galactic $R_V=3.1$.  Early type elliptical galaxies typically have low $A(V)$ \citep[see e.g.,][who found $A(V)\lesssim0.35$ for dust lanes in ellipticals]{goudfrooij1994}.

\citet{nadeau} pointed out that gravitationally lensed quasars could be used to measure the extinction curves of higher redshift galaxies.   \citet{falco1999}  explored a large sample of 23 lensing galaxies using this method and found that only seven were consistent with no extinction.  This method has also been applied to single systems by e.g. \citet{jaunsen,toft,motta,wucknitz,munoz,wisotzki,goicoechea} and shows varying extinction properties between different lensing systems.

Here we present a systematic study of the extinction curves of gravitational lenses based on a survey of $10$ lens systems.  We have made a dedicated effort to minimize the number of unknowns and effects that can mimic extinction.  We have broad wavelength coverage in nine different optical and NIR broad bands. An effort was made to minimize the time between the observations for each system in the different bands to minimize the effect of intrinsic quasar variability and microlensing.  All our systems have spectroscopically determined redshifts for both the quasar and the lensing galaxy.  Finally, our systems span the range of $z=0.04$--$1.01$ giving us the possibility to explore possible evolution with redshift.

The outline of this paper is as follows:  In \S~\ref{sec:method} we describe the details of the employed method and discuss different sources of systematics and of random errors which may affect our results.  We also present the results of simulations which explore the effects of these errors on data sets similar to those we obtain in the survey.  In \S~\ref{sec:data} we present the data and the data reduction of the ESO VLT survey for the 10 lensing systems.  We present the results of our analysis of each individual system in \S~\ref{sec:res_ind} and the analysis of the full sample in \S~\ref{sec:res_full}.  Finally we summarize our results in \S~\ref{sec:summary}.

\section{Method and simulations}
\label{sec:method}
In this section we introduce the method and explore the different sources of systematic and random errors through simulations.  In particular, we study the effects of achromatic microlensing and of extinction along both lines of sight, and use simulations to explore the conditions, under which it will be possible to recover and distinguish between different extinction laws. 

\subsection{Lensing}
\label{sec:lensing}
\begin{figure}
\epsscale{0.85}
\plotone{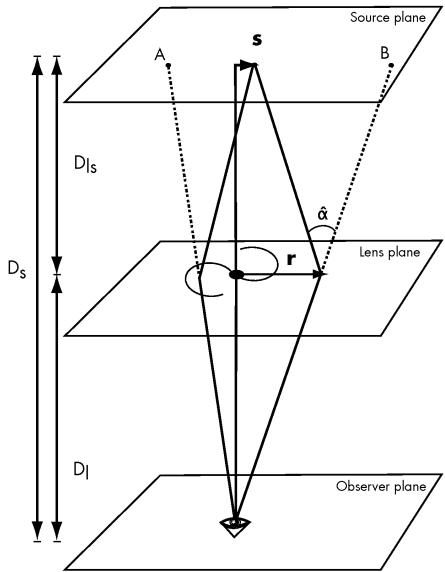}
\caption{The lensing setup.  $D_l$ is the distance to the lens, $D_s$ the distance to the source, $D_{ls}$ the distance from the lens to the source, $\hat{\alpha}$ is the deflection angle (see eq.~\ref{eq:lensalpha}), $\mathbf{s}$ is the position vector of the source in the source plane, and $\mathbf{r}$ is the position vector in the lens plane.  The figure shows a doubly imaged source, where one line of sight goes through the lensing galaxy and the other passes outside it. \label{fig:lenssetup}}
\end{figure}

Gravitational lensing is the deflection of light rays due to the gravitational field of the matter distribution through which the light passes.  For a geometrically thin lens, i.e., where the depth of the lens is small compared to the distance between the lens and observer, $D_l$, and the lens and the light source, $D_{ls}$, the deflection angle is given by
\begin{equation}
\mathbf{\hat{\alpha}}(\mathbf{r}) = \frac{4 G}{c^2} \int d^2 r^{'} \Sigma(\mathbf{r^{'}}) \frac{\mathbf{r-r^{'}}}{|\mathbf{r-r^{'}}|^2}
\label{eq:lensalpha}
\end{equation}
where $G$ is the gravitational constant, $c$ is the speed of light, $\Sigma(\mathbf{r})$ is the surface mass density of the lensing mass and $\mathbf{r}$ is the impact vector of the light ray (see Figure~\ref{fig:lenssetup} for a sketch of the lensing setup).  For a position vector $\mathbf{s}$ in the source plane one will see images at locations $\mathbf{r}$ in the lens plane which satisfy the lens equation
\begin{equation}
\mathbf{s} = \frac{D_s}{D_l}\mathbf{r} - D_{ls} \mathbf{\hat{\alpha}}(\mathbf{r})
\label{eq:lens}
\end{equation}
where $D_{ls}$ is the angular diameter distance between the lens and the source and where we have assumed that the size of the lens is small compared to $D_l, D_s$ and $D_{ls}$.
  When eq. (\ref{eq:lens}) has more than one solution we see multiple images of the source in the lens plane which are in general located at different distances from the center of the mass distribution.  In the case of distant quasars being lensed by foreground galaxies, the condition that the lens size be small compared to $D_l,D_s$ and $D_{ls}$ is fulfilled.

The multiple images of lensed quasars act as `standard candles' shining through different parts of the lensing galaxy and can therefore be used to study its extinction curve \citep[as first pointed out by][]{nadeau}.  As lensing is achromatic in nature, the flux ratio of any two lensed images should be independent of wavelength. If, however, one of the images shines through a dusty part of the galaxy and the other image does not, the first image  will appear red compared to the other. By mapping the flux ratio as a function of wavelength one can in principle directly trace the differential extinction curve between the two images, without making any assumptions about the intrinsic spectrum of the quasar (as it cancels in the calculation of the flux ratios).  Depending on the number of images, it will be possible to obtain differential extinction curves for several paths through the lensing galaxy.

As the light rays travel along different paths for multiple images, their travel times do in general differ, introducing a time delay between different images.  If the quasar is variable, the variability will show up at different times in the different images which can lead to inaccurate estimates of the dust extinction.  Ideally one would like to measure each image with a time separation according to the time delay to correct for this effect.  Time delays are, however, difficult to measure and measurements exist for only a few lenses.  Alternatively one can observe simultaneously in all the observing bands.  This would mean that any achromatic variability would cancel out when comparing the images (but would lead to biased estimates of the intrinsic brightness ratio of the images).  Simultaneous observations also have the additional benefit that the effects of achromatic microlensing will, to first order, only affect the intrinsic ratio estimate and not the shape of the extinction curve, which would in general not be the case if the images were observed according to a time schedule given by the time delay.  The effects of microlensing are the greatest potential source of systematic error in our extinction curve analysis and are addressed in greater detail in \S~\ref{sec:micro}.

\subsection{Extinction}
As lensing is an achromatic process we would expect the magnitude difference in all bands to be constant for each pair of lensed images in the absence of extinction.  Extinction reduces the brightness of the measured images by a different amount for each band (and image) depending on the amount and properties of the dust along the line of sight to the images.  It is this difference which gives rise to the extinction curve as a function of wavelength.  As both the images might be affected by extinction what one is really measuring is the differential extinction between the pair of images.  The extinction affects each measured data point as:
\beq
m(\lambda)=\hat{m}(\lambda) + A(\lambda),
\eeq
where $m(\lambda)$ is the measured magnitude of the image, $\hat{m}(\lambda)$ is the intrinsic magnitude of the image and $A(\lambda)$ is the extinction at wavelength $\lambda$.  When comparing images A and B one therefore gets:
\begin{eqnarray}
\Delta m (\lambda ) & \equiv & m^B(\lambda) - m^A(\lambda)\\
                    & =      & (\hat{m}^B-\hat{m}^A) + A^B(\lambda)-A^A(\lambda) \nonumber\\
                    & \equiv & \Delta\hat{m}+ A^{diff}(\lambda),\nonumber
\label{eq:fiteq}
\end{eqnarray}
where $\Delta\hat{m}\equiv\hat{m}^B-\hat{m}^A$ is the intrinsic magnitude difference which does not depend on wavelength and $A^{diff}(\lambda)\equiv A^B(\lambda)-A^A(\lambda)$ is the effective differential extinction law as a function of the wavelength.

We consider three different extinction laws and assume that the extinction of one of the images dominates the other (see further discussion on extinction in both images in \S~\ref{sec:extboth}).  The first extinction law we consider is the empirical \cardelli~as parametrized by \citet{cardelli}:
\begin{eqnarray}
\label{eq:car}
A(\lambda)& = & E(B-V) \left[R_V a(x) + b(x)\right] \\
          & = &  A(V) \left[ a(x) + \frac{1}{R_V} b(x)\right],\nonumber
\end{eqnarray}
where $A(\lambda)$ is the total extinction at wavelength $\lambda$, $E(B-V)=A(B)-A(V)$ is the color excess, $R_V=A(V)/E(B-V)$ is the ratio of total to selective extinction, $a(x)$ and $b(x)$ are polynomials and $x=\lambda^{-1}$.  We also consider an extinction law which is linear in inverse wavelength which is characteristic for the extinction in the SMC:
\beq
\label{eq:lambda}
A(\lambda) = A(V) \left(\frac{ \lambda}{5500\AA}\right)^{-1}.
\eeq
Finally we extend the linear law to a power law:
\beq
\label{eq:alpha}
A(\lambda) =  A(V) \left(\frac{ \lambda}{5500\AA}\right)^{-\alpha}.
\eeq

To fit data points to the extinction laws one first shifts the wavelength of the measured bands to the rest frame of the lensing galaxy, i.e. $\lambda_j=\lambda_j^O/(1+z_l)$ where $\lambda_j^O$ is the observed wavelength in band $j$ and $z_l$ is the redshift of the lensing galaxy.   For each image pair one can then calculate the magnitude difference of the images in each measured band:
\beq
m^B(\lambda_j)-m^A(\lambda_j) = -2.5 \log_{10}\left(\frac{f_j^B}{f_j^A}\right),
\eeq
where $f_j^B/f_j^A$ is the flux ratio between images labeled B and A at $\lambda_j$.   One can then perform fits for eq. (\ref{eq:fiteq}) where one replaces $A^{diff}(V)$ with $A(V)$:
\beq
\label{eq:fit}
m^B(\lambda)-m^A(\lambda) =A(\lambda) + \Delta\hat{m},
\eeq
where $A(\lambda)$ is one of the extinction laws described in eqs.~(\ref{eq:car}), (\ref{eq:lambda}) and (\ref{eq:alpha}).  If radio measurements exist for the flux density of the images, one can use them to constrain the intrinsic magnitude difference, $\Delta\hat{m}$, as radio measurements are not affected by extinction.

\subsection{Extinction along both lines of sight}
\label{sec:extboth}
As the method we use measures differential extinction curves, we wish to investigate the systematics of extinction along both lines of sight in our results.  We therefore study the effects of a \cardelli~along both lines of sight but with different values of $R_V$.   When both images suffer extinction we expect to get an effective extinction law which may have different properties to those of either line of sight.  In the general case, when $E^B(B-V)\ne E^A(B-V)$, the difference in extinction suffered by image A vs. image B will be given by an effective \cardelli:
\begin{eqnarray}
A^{diff}(\lambda) & \equiv  & A^B(\lambda) - A^A(\lambda) \\
        & =       & E^B (R_V^B a(\lambda^{-1}) + b(\lambda^{-1})) -\label{eq:dif2}\\
 & &E^A (R_V^A a(\lambda^{-1}) + b(\lambda^{-1})) \nonumber\\
        & =       & \left(E^B-E^A\right) \left(\frac{E^B R_V^B - E^A R_V^A}{E^B-E^A} a(\lambda^{-1}) + b(\lambda^{-1})\right) \label{eq:dif3} \\
        & \equiv & \left(E^B-E^A\right) \left(R_V^{diff} a(\lambda^{-1}) + b(\lambda^{-1})\right) \nonumber 
\end{eqnarray}
where we have written $E(B-V) = E$ for simplicity, explicitly used the assumption $E^B\ne E^A$ in the step from equation (\ref{eq:dif2}) and (\ref{eq:dif3}) and $R_V^{diff}\equiv(E^BR_V^B-E^AR_V^A)/(E^B-E^A)$ is the effective $R_V$ we measure.  For completeness we note that in the special case of $E^A=E^B$ the resulting effective extinction curve is not given by the \cardelli~parametrization.  If we take $E^B>E^A$, we find that the ratio of $R_V^{diff}$ to the $R_V^B$ (which we wish to measure) is given by:
\begin{eqnarray}
\frac{R_V^{diff}}{R_V^B} & = & \frac{E^B - E^A R_V^A/R_V^B}{ E^B - E^A} \\
                    & = & 1 + \frac{E^A}{E^B-E^A}\left( 1 - \frac{R_V^A}{R_V^B}\right) \nonumber \\
                    & \equiv & 1 + \eta \nonumber
\label{eq:rdiff}
\end{eqnarray}
so the error introduced in our estimate due to the non-zero extinction of image A is 
\beq
\eta = \frac{E^A/E^B}{1-E^A/E^B}\left( 1 - \frac{R_V^A}{R_V^B}\right).
\label{eq:eta}
\eeq
We note that if $R_V^A > R_V^B$, the inferred value of $R_V$ for image B will be lowered, and vice versa, and that, in theory, any value of $R_V$ can be obtained. 
\begin{figure}
\epsscale{1.0}
\plotone{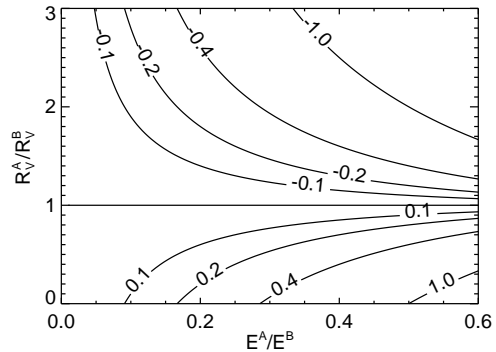}
\caption{Contour plot of $\eta$ as defined in eq. (\ref{eq:eta}) which measures the relative error of the effective $R_V^{diff}$ to $R_V^B$ as a function of $E^A/E^B$ and $R_V^A/R_V^B$ for extinction along both lines of sight. \label{fig:eta}}
\end{figure} 
A contour plot of $\eta$ can be seen in Figure~\ref{fig:eta}.  If we take the most extreme values of the Milky Way ($2\lesssim R_V\lesssim 6$) we see that the bracket in equation (\ref{eq:eta}) can realistically range from $(1-6/2)=-2$ to $(1-2/6)=2/3$.  In the worst case scenario, when $R_V^B\ll R_V^A$, we therefore need the ratio of $E^A/(E^B-E^A)$ to be half that of the desired accuracy, i.e., for a desired accuracy of $10\%$ in $R_V$ we need $E^A/(E^B-E^A)\le0.05$.  For more realistic cases of $R_V=3$ for image B and $R_V=4$ for image A we only need $E^A/(E^B-E^A) \le 0.3$.  In general, for an accuracy of $|\eta|\le\eta_0$, we need:
\beq
\frac{E^A}{E^B} \le \frac{\eta_0/|1-R_V^A/R_V^B|}{1+ \eta_0/|1-R_V^A/R_V^B|}.
\label{eq:accuracy}
\eeq

Finally we note for completeness that a linear extinction in both images trivially produces a linear differential extinction whereas a power law extinction along both lines of sight does not in general produce a power law for the differential extinction.

\citet{mcgough} did a similar study of the effect of non-zero extinction along both lines of sight, where they also found that, in theory, any value of $R_V$ can be obtained.  They suggest that cases where only one of the images is lightly reddened or the dust properties are the same for both sight lines are likely rare and hard to confirm.  We point out that it is not crucial that one of the images have no or little extinction in absolute terms, but only relative to the image we are comparing it to, and that in practice, one of the images often shows less extinction than the others.  When dealing with multiply imaged quasars, in particular for doubly imaged systems, one of the images is often situated at a greater distance from the lens galaxy than the others and therefore may be less affected by extinction.  In some cases this lack of extinction in one of the components can be confirmed by studying the images in the X-rays (K. Pedersen et al., 2006, in preparation). 

\subsection{Microlensing}
\label{sec:micro}
Microlensing, lensing by stars or other compact objects in the lens galaxy, can also affect our data and in particular it can affect the continuum part of the emission.  This is because, according to standard quasar models \citep{krolik}, the regions giving rise to the continuum and emission lines are of different size and therefore affected differently by microlensing which is more effective on small scales.  As the emission lines arise from regions several orders of magnitude larger than the region emitting the continuum, the microlensing acts strongest on the continuum emission but should be nearly absent for emission lines.

\begin{figure}
\epsscale{1.0}
\plotone{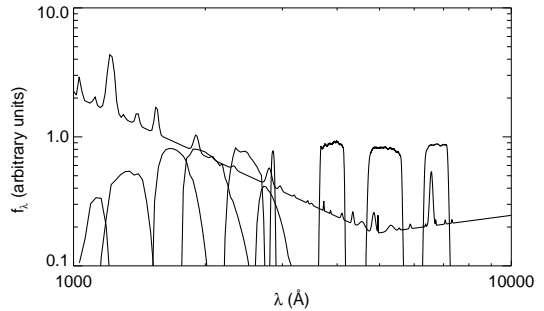}
\caption{The composite quasar spectra modeled as a broken power law with spectral lines with equivalent width $W \geq 1$ $\AA$ added on as Gaussians.  The superimposed transmission curves correspond to the FORS1 and ISAAC $U, B, V, R, I, z$ (Gunn) $, z$ (special), $J_s, H$ and $K_s$ bands.  The transmission curves have been shifted to show which part of the spectrum they correspond to for a quasar of $z=2.2$. \label{fig:quasar_spectra}}
\end{figure}

To allow for possible corrections due to this effect we calculate for each quasar the ratio of the spectral line emission to the total emission in each band (this varies for the systems as the quasars are at different redshifts).  We use a composite quasar spectrum for our calculations as derived by \citet{berk} using 2200 spectra from the Sloan Digital Sky Survey.  In accordance with their results we model the continuum as a broken power law ($f_\lambda\propto\lambda^{\alpha_\lambda}$) with $\alpha_\lambda=-1.56$ for $\lambda\leq4850$~$\AA$ and $\alpha_\lambda=0.45$ for $\lambda>4850$~$\AA$.  We take into account spectral lines with equivalent width $W\geq1$~$\AA$ and model them as Gaussians which we add to the continuum to get the final template spectrum.  Using this standard quasar template we calculate the ratio of the flux coming from the continuum compared to the total emission for each measurement band shifted to the corresponding band at the redshift of the lensing galaxy using the transmission curves of the corresponding filter (see Figure~\ref{fig:quasar_spectra}).

We add to our fit an effect from an achromatic microlensing signal which affects the different bands proportionally to the ratio of the continuum to the total emission (consisting of the continuum with the added spectral lines).  Achromatic microlensing affects the fluxes of the images as $f_j\rightarrow f_j(1+sr_j)$  where $r_j$ is the ratio of the continuum emission to the total emission in band $j$ and $s$ is a microlensing parameter giving the strength of the microlensing signal (which is a constant in the achromatic case).  The measured magnitude is therefore:
\begin{eqnarray}
\label{eq:micro}
m(\lambda_j) &= &-2.5 \log_{10} \left(f_j (1 + s r_j)\right) + A(\lambda_j) \\
             &= &\hat{m}(\lambda_j) - 2.5 \log_{10}(1+s r_j) + A(\lambda_j).\nonumber
\end{eqnarray}
The additional term in eq. (\ref{eq:micro}) modifies eq. (\ref{eq:fit}) to:
\begin{eqnarray}
\label{eq:extmicro}
(m^B - m^A)(\lambda_j) &= &A(\lambda_j) + \Delta\hat{m} - 2.5 \log_{10}\left(\frac{1+s^B r_j}{1+s^A r_j}\right) \\
&\approx & A(\lambda_j) + \Delta\hat{m} - 2.5 \frac{r_j(s^B-s^A)}{\ln(10)} \nonumber\\
&\equiv  &A(\lambda_j) + \Delta\hat{m} + s r_j,\nonumber
\end{eqnarray}
where $s\equiv2.5(s^A-s^B)/\ln(10)$ and the approximation is made to reduce the number of parameters in the fit.  For the approximation in eq. (\ref{eq:extmicro}) to be valid we need $r_j|s^B-s^A|\lesssim1$.  The value of $r_j$ can in theory lie between 0 and 1 but in our case lies between 0.8 and 1, so therefore we need $|s^B-s^A|\lesssim1$ in practice.  In physical terms this is roughly equivalent to the condition that the change in magnitude difference between the two images due to microlensing should be less than $1$ mag.  Most microlensing studies show magnitude changes of less than $0.5$ mag so therefore we do not expect this approximation to affect our results.

Microlensing can also introduce a smooth chromatic signal similar to an extinction signal if the source has different colors in different locations.  Our methods however do not take this into account as our broad band photometry does not contain enough information to disentangle such an effect from extinction.  To separate the two it is necessary to study the spectra of the object itself (and not a mean quasar spectrum) along the lines of \citet{wucknitz}.  

\subsection{Monte Carlo Analysis}

We estimate the errors of the parameters of the fitted extinction curves using
Monte Carlo simulations. We generate a thousand realizations of the data set by
allowing each data point to vary with a Gaussian distribution centered on the measured point and with a variance corresponding to the error estimate of the data point.  We then apply the extinction law fits separately to each data set.  If the fits are good, i.e., there is no degeneracy or systematic errors, the resulting fitted parameters should also follow a Gaussian distribution.  We quote the median of the parameter distribution as our `best fit parameter' and the quoted error is the standard deviation of the parameter distribution.  For the \cardelli~we constrain $0\leq R_V\leq7$ unless otherwise noted.  This method is applied to both the simulations in \S~\ref{sec:sim} and the real data sets in \S~\ref{sec:results}.

\subsection{Simulated Data}
\label{sec:sim}
To test the ability of our analysis to recover and distinguish between different extinction curves, and the effects of microlensing and noise, we apply our method to simulated data.  Our simulated data consist of data in nine bands, $UBVRIzJsHK$, with an applied signal corresponding to extinction, noise and microlensing.  We do runs for different kinds of the three extinction laws, with or without noise, and with or without microlensing effects.

\subsubsection{Pure extinction in one image}
\label{sec:pureext}
The first set of runs we do are pure extinction in one image with a $0.05$ mag $1\sigma$ uncertainty in each data point.  The purpose of those runs is twofold, first to test our routines and secondly to see whether there is a significant difference between the goodness of fit for the different extinction laws.  We find that in all cases our routines converge to the given initial parameters.  In addition we find that the ability to recognize one extinction law from the other depends on the strength of the extinction and the redshift of the simulated data.  This is not surprising as the three fitted extinction laws behave very similarly for $\lambda\ge2500$~$\AA$ and for $z=0$ all our bands lie above that limit.  The importance of the strength of the extinction is also easy to understand, as the ability to detect any difference between extinction laws will be overwhelmed by the photometric errors for very weak extinction.

\begin{figure}
\epsscale{1.0}
\plotone{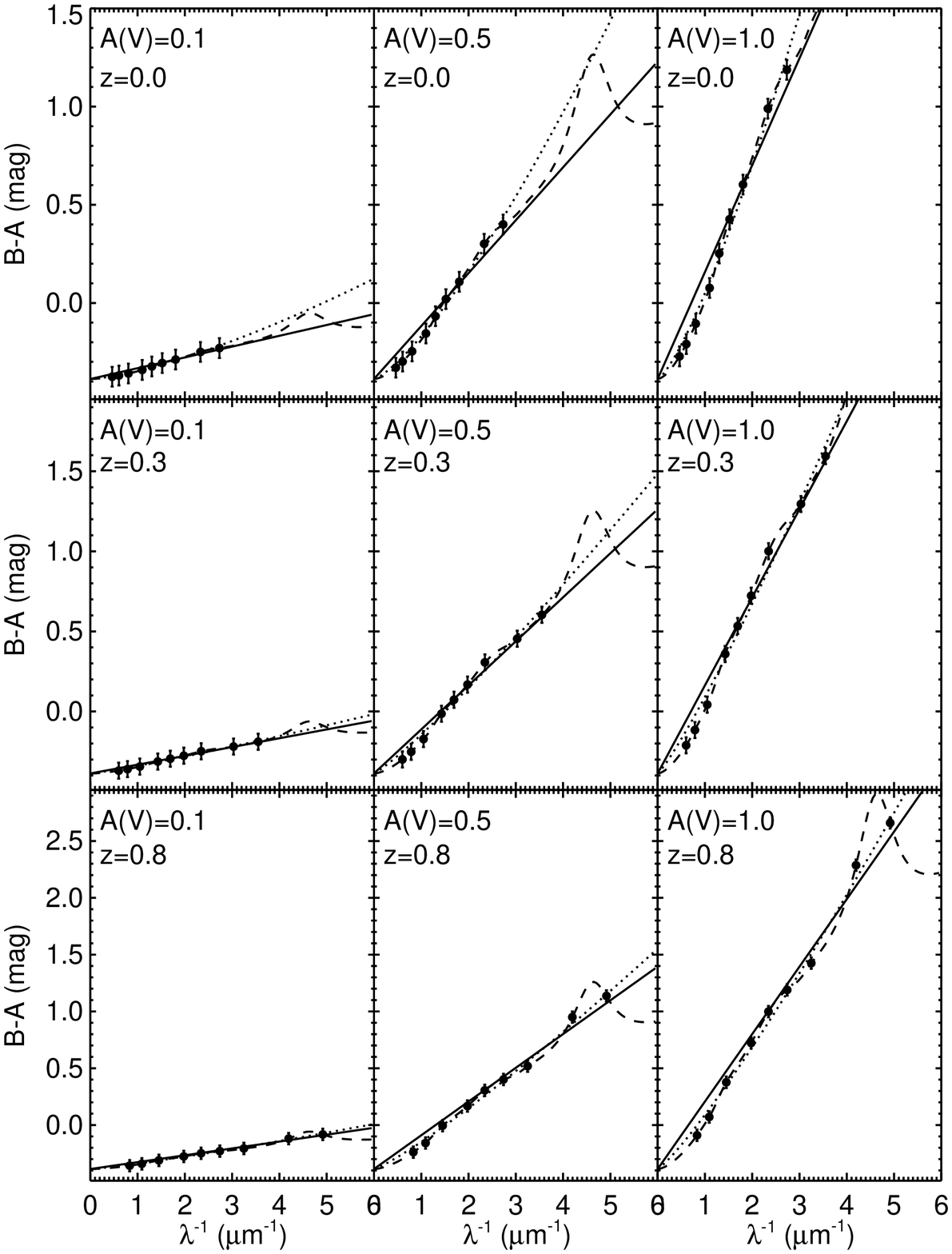}
\caption{ A sample plot for simulations showing three data sets at redshift of $z=0.0,0.3$ and $0.8$.  The extinction law applied is the \cardelli~with $R_V=3$ and $A(V)=0.1, 0.5, 1.0$ and the three fits shown are the \cardelli~(dashed line, eq. (\ref{eq:car})), power law (dotted line, eq. (\ref{eq:alpha})) and linear law (solid line, (\ref{eq:lambda})).\label{fig:sim_strength}}
\end{figure}

\begin{figure}
\epsscale{1.0}
\plotone{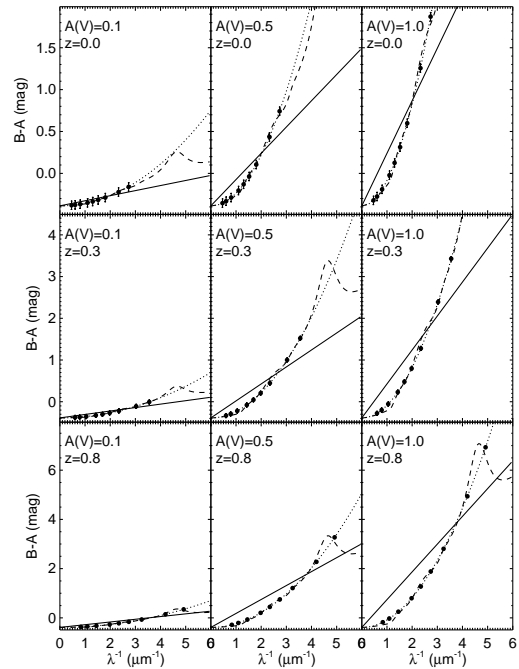}
\caption{ A sample plot for simulations showing three data sets at redshift of $z=0.0,0.3$ and $0.8$.  The extinction law applied is the power law with $\alpha=2$ and $A(V)=0.1, 0.5, 1.0$ and the three fits shown are the \cardelli~(dashed line, eq. (\ref{eq:car})), power law (dotted line, eq. (\ref{eq:alpha})) and linear law (solid line, (\ref{eq:lambda})).\label{fig:sim_strength_alpha}}
\end{figure}

At $z=0$  for a \cardelli~input the reduced chi-squared, $\chi^2_{\nu}$, for a \cardelli~is somewhat lower than for the other extinction laws ($\approx1.0$ for the \cardelli, $ \approx 1.0$--$1.6$ for the power law, $\approx 1.0$--$6$ for the linear law).  The difference depends both on the strength of the extinction, $A(V)$ (with $A(V)\lesssim0.1$ resulting in equal goodness of fits), and on the value of $R_V$ with more extreme values ($R_V=1$, $R_V=6$) giving a greater difference than $R_V=3.0$ (see Table~\ref{tab:chisim} for representative values and Figures \ref{fig:sim_strength} and \ref{fig:sim_strength_alpha} for representative plots).  When the input extinction law is a linear law, $\lambda^{-1}$, the power law yields the same goodness of fit ($\chi^2_{\nu}\approx1.0$) but the \cardelli~gives a slightly higher value ($\approx1.0$--$1.9$).  Finally, when the input extinction law is a power law with power index $\alpha=2$, there is a significant difference in the goodness of fit with the power law fit giving $\chi^2_{\nu}\approx1.0$ but the \cardelli~giving $\chi^2_{\nu}\approx1.0$--$1.7$ and $\lambda^{-1}$ giving $\chi^2_{\nu}\approx1.0$--$6.5$.  Here the difference in the goodness of fit for the \cardelli~and $\lambda^{-1}$ depends on the value of $A(V)$, with lower $A(V)$ (i.e., less extinction) giving lower $\chi^2_{\nu}$ for the two extinction laws.

\begin{deluxetable*}{lllllrrrrrr} 
\tablecolumns{11} 
\tablewidth{0pc} 
\tablecaption{Goodness of fits from simulations} 
\tablehead{
\multicolumn{4}{c}{Input parameters}    & \colhead{}   & \multicolumn{6}{c}{Output $\chi^2_\nu$ from the different fits}\\ 
\cline{1-4} \cline{6-11}\\
  \colhead{$z$}& \colhead{Type} & \colhead{$A(V)$} & \colhead{Parameter} & \colhead{}  & \colhead{(1)}   & \colhead{ (2)} & \colhead{ (3)}  & \colhead{(4)}&\colhead{(5)}& \colhead{(6) }}

\startdata 
0.0 & \cardellitab &0.5& $R_V=1.0$ & & 0.94 &0.93	 &1.6 &1.6 &5.9 &3.5\\  
0.3 & \cardellitab &0.5& $R_V=1.0$ & & 0.97 &0.96	 &1.9 &1.6 &8.5 &4.2\\  
0.8 & \cardellitab &0.5& $R_V=1.0$ & &  0.95&	0.94 &3.4 &2.9 &15 &6.8\\   

0.0 & \cardellitab &1.0& $R_V=3.0$ & &  0.95&	0.94  & 1.4& 1.1& 2.5& 1.1\\  
0.3 & \cardellitab &1.0& $R_V=3.0$ & &  0.96&0.96	& 1.9& 1.2& 2.3& 1.5\\  
0.8 & \cardellitab &1.0& $R_V=3.0$ & &  0.96& 0.94	 & 1.8& 1.7& 2.9&1.6\\   

0.0 & \cardellitab &0.5& $R_V=3.0$ & & 0.95 &	 0.94 &1.1 &1.0 &1.5 &1.0\\  
0.3 & \cardellitab &0.5& $R_V=3.0$ & & 0.95 &0.95	& 1.2& 1.0&1.4 &1.1 \\  
0.8 & \cardellitab &0.5& $R_V=3.0$ & &  0.96&	0.94 &1.2 &1.2 &1.6 &1.2\\   

0.0 & \cardellitab &0.1& $R_V=3.0$ & & 0.97  &0.97 &0.95 &0.94 &0.98 &0.92\\  
0.3 & \cardellitab &0.1& $R_V=3.0$ & & 0.95 &0.94	 &0.97 &0.98 &0.97 &0.96\\  
0.8 & \cardellitab &0.1& $R_V=3.0$ & & 0.95 &0.95&	 0.97& 0.98& 0.99& 0.96\\   

0.0 & \cardellitab &0.5& $R_V=6.0$ & & 0.96 &0.96 &	 1.3& 1.0& 1.2& 1.3\\  
0.3 & \cardellitab &0.5& $R_V=6.0$ & & 0.96 &0.96	 &1.7 &1.3 &1.9 &1.8\\  
0.8 & \cardellitab &0.5& $R_V=6.0$ & & 0.95 &0.94&	 1.6& 1.3& 2.7& 1.9\\   

0.0 & Power law &1.0& $\alpha = 2$ & & 1.6& 1.7& 0.96&0.95 &6.5& 3.5\\  
0.3 & Power law &1.0& $\alpha = 2$ & & 2.3& 2.1&0.95 & 0.95&11 &5.7  \\  
0.8 & Power law &1.0& $\alpha = 2$ & & 4.7& 4.5& 0.95& 0.93&21&11 \\

0.0 & Power law &0.5& $\alpha = 2$ & &1.1 &1.2 &0.97 &0.96 &3.4 &1.9 \\  
0.3 & Power law &0.5& $\alpha = 2$ & & 1.4& 1.3& 0.96& 0.95& 5.5& 3.0 \\  
0.8 & Power law &0.5& $\alpha = 2$ & & 2.4& 2.3& 0.94& 0.94& 11& 5.5\\   

0.0 & Power law &0.3& $\alpha = 2$ & & 1.0& 1.0& 0.97& 0.97& 2.2& 1.4\\  
0.3 & Power law &0.3& $\alpha = 2$ & & 1.1& 1.1& 0.96& 0.95& 3.4 &1.9 \\  
0.8 & Power law &0.3& $\alpha = 2$ & &1.6 &1.6 &0.95 &0.94 &6.4 &3.4 \\   

0.0 & Power law &0.1& $\alpha = 2$ & &0.95 &0.95 &0.94 & 0.94&1.2&1.0 \\  
0.3 & Power law &0.1& $\alpha = 2$ & & 0.97& 0.96& 0.95& 0.96& 1.4&1.1  \\  
0.8 & Power law &0.1& $\alpha = 2$ & & 1.0&1.0 &0.94 &1.0 &2.3&1.4 \\   

0.0 & Linear &1.0& $\alpha = 1$ & &1.9 &1.1 & 0.96&0.94 &0.96&0.96 \\  
0.3 & Linear &1.0& $\alpha = 1$ & & 2.2& 1.2& 0.95& 0.95& 0.96&  0.95\\  
0.8 & Linear &1.0& $\alpha = 1$ & & 1.9& 1.5& 0.95& 0.95&0.96&0.95 \\   

0.0 & Linear &0.5  & $\alpha = 1$ & & 1.3& 1.0& 0.96& 0.96& 0.96& 0.96\\  
0.3 & Linear &0.5  & $\alpha = 1$ && 1.4& 1.0 &0.96 & 0.94 & 0.96 & 0.95 \\  
0.8 & Linear &0.5  & $\alpha = 1$ && 1.3& 1.1& 0.94& 0.93& 0.96& 0.94\\   

0.0 & Linear &0.3  & $\alpha = 1$ && 1.1& 0.96& 0.96& 0.95& 0.96& 0.96\\  
0.3 & Linear &0.3  & $\alpha = 1$ && 1.1& 0.97& 0.96& 0.95 & 0.95& 0.95\\  
0.8 & Linear &0.3  & $\alpha = 1$ && 1.1& 1.0&0.94 &0.92 &0.95 &0.94 \\   

0.0 & Linear &0.1& $\alpha = 1$ & & 0.98 & 0.95& 0.95&0.95 &0.95& 0.95\\  
0.3 & Linear &0.1& $\alpha = 1$ & &0.98 &0.97 &0.96 &0.99 &0.97 &0.96  \\  
0.8 & Linear &0.1& $\alpha = 1$ & & 0.97&0.95 &0.95 & 0.99&0.96&0.95 \\   

\enddata 
\tablecomments{Table of $\chi^2_{\nu}$ for representative values of the simulated data.  The output columns correspond to \cardelli~with $\Delta\hat{m}$ fixed (1) and free (2), power law with  $\Delta\hat{m}$ fixed (3) and free (4), and linear law with  $\Delta\hat{m}$ fixed (5) and free (6).}
\label{tab:chisim}
\end{deluxetable*}

We also run simulations for two non-zero values of the redshift, at $z=0.30$ and $z=0.80$.  At these higher redshifts the difference in the extinction laws becomes more prominent which is reflected in the goodness of the fits (see Table~\ref{tab:chisim} and Figs.~\ref{fig:sim_strength} and \ref{fig:sim_strength_alpha}) making them in general easier to distinguish from one another.  This is because the extinction laws are all very similar for $\lambda^{-1}\lesssim3$~$\mu m^{-1}$, and for redshift of $z=0$ our lowest wavelength band, the $U$-band, lies below this limit.  At the higher redshifts, the lowest wavelengths move into the UV, which is more sensitive for differences in the extinction laws.  For very weak extinction ($A(V)\lesssim0.1$) it is still the case that the different types of extinction laws become hard to separate, as the error bars can dominate the extinction signal.

From our simulations we can deduce that given a strong enough extinction the different kinds of extinction laws should  be recognizable from each other.  As we can see from Table \ref{tab:chisim}, the needed strength is dependent on the redshift of the lens, with more nearby lenses requiring stronger extinction to distinguish the different extinction laws.  In general, we find that for $A(V)\ga0.3$ we can distinguish the different extinction laws, in particular if we have constraints on $\Delta\hat{m}$.  The parameters of the power law fit tend to be more poorly constrained than those of the other fits.  This is due to the added degree of freedom which can result in several sets of parameters giving similar $\chi^2_\nu$.  We also see that if we can constrain the intrinsic magnitude ratio, the different extinction laws become easier to distinguish.

\subsubsection{Extinction in both images}
\label{sec:extboth_sim}
We first simulate data where we apply the same extinction law to both images (i.e., $R_V$ and $\alpha$ are constant respectively) but at different strengths.  Here we find that unless $A^A(V)\approx A^B(V)$ (i.e., in effect a very weak differential extinction), the fits converge to the starting parameters and give the shape of the `real' extinction curves.

\begin{figure}
\epsscale{0.85}
\plotone{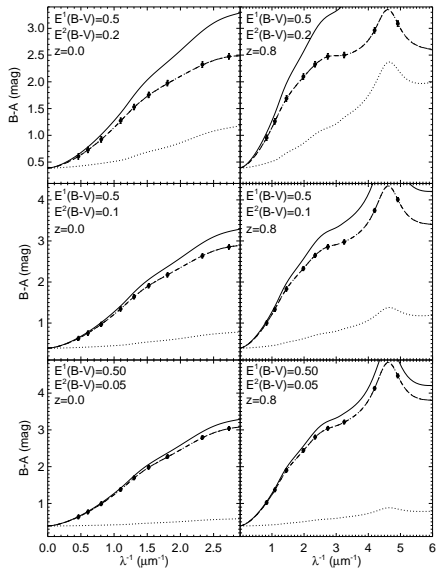}
\caption{A sample plot for simulations with Galactic type extinction in both images at two different redshifts ($z=0.0$ and $0.8$).  The solid and the dotted lines are the input extinction laws and the dashed line is the effective differential extinction between the two images.  We see that when the extinction along one line of sight is much stronger than the other, then the effective extinction law approaches the extinction of the more strongly extinguished image.\label{fig:sim_both}}
\end{figure}

To see what effect extinction in both images would have on a data set, we simulate data having different kinds of \cardelli s, to find within which accuracy the input parameters of the more strongly extinguished line of sight are found (see Fig. \ref{fig:sim_both}).  Again we find, that if the differential extinction between the images, $E^B(B-V)-E^A(B-V)$, is low, then the fits are not well constrained and the resulting parameters represent the extinction of neither line of sight.  However, if one image is significantly more extinguished than the other then the parameters of the fit converge to the parameters of that line of sight as we expect (see discussion in \S~\ref{sec:extboth}).  It is not crucial that one image be non-extinguished, but it does need to have significantly lower extinction than the other line of sight.  We find that for the fits to be within one sigma of the real parameters for the line of sight for the stronger extinction we need roughly $E(B-V)_{weaker}/E(B-V)_{stronger}\lesssim0.2$ when the $R_V$ values lie in the range of $2-4$.  This is consistent with our results from \S \ref{sec:extboth} (in particular, see eq. (\ref{eq:eta}) and (\ref{eq:accuracy})).

\subsubsection{The effects of noise}
\begin{figure}
\epsscale{0.85}
\plotone{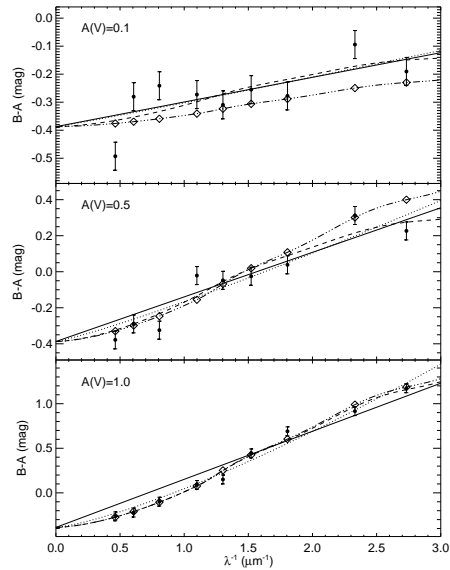}
\caption{A sample plot for simulations with noise showing three data sets with $A(V)=0.1, 0.5$ and $1.0$.  The diamonds and the dash-dotted line show the noiseless input extinction law (\cardelli~with $R_V=3.0$), and the solid dots show the data points with random noise applied.  The three fits shown are the \cardelli~(dashed line, eq. (\ref{eq:car})), power law (dotted line, eq. (\ref{eq:alpha})) and linear law (solid line, (\ref{eq:lambda})).  We see that for low extinction ($A(V)\lesssim 0.1$) the noise is of the order of the amount of the extinction signal, and therefore significantly affects the quality of the fits. \label{fig:sim_noise}}
\end{figure}

When dealing with real data we can expect our data sets to be contaminated by various sources of noise.  To see how this may affect our analysis we generate data sets with artificial random noise (see Fig. \ref{fig:sim_noise} for representative plots).  We generate the noise as normally distributed random numbers with a mean of $0$ and standard deviation of $0.05$ magnitudes (which is an estimate of the lowest noise expected from deconvolved ground based data such as our data set).  We still get qualitatively the same results, i.e., we get the lowest $\chi^2_\nu$ for the Galactic type extinction fit to the Galactic extinction law data (provided the extinction is strong enough), and the fitted parameters agree with the input parameters, within the uncertainty.  The $\chi^2_{\nu}$ values for the fits are increased and the uncertainty on the parameters for the noisy data is generally larger.  If the noise becomes too large, i.e. of the order of the extinction effects, the fits become badly constrained.  This emphasizes the need for a strong extinction signal to outweigh random noise effects when analyzing the real data sets.

\subsubsection{Achromatic microlensing}

\begin{figure}
\epsscale{0.85}
\plotone{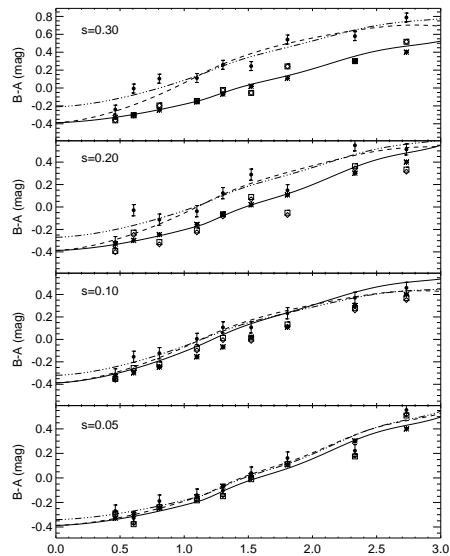}
\caption{A sample plot for simulations with an achromatic microlensing signal $s=0.05,0.10,0.20,0.30$ and noise.  The asterisks are the noise free input data points, the boxes are the input data points with noise, the solid circles are the microlensed data points we fit and the diamonds are the microlensing corrected points from the fit.  The dashed and dash-dotted lines are the best fitting \cardelli~with $\Delta \hat{m}$ free and fixed through the microlensed data points and he solid line shows the best fitting \cardelli~with $\Delta\hat{m}$ free with a microlensing correction.  The microlensing signal starts dominating the noise when the average distance from the asterisks to the boxes becomes smaller than the average distance from the boxes to the filled circles which happens around $s=0.05$.\label{fig:sim_micro}}
\end{figure}
\begin{deluxetable*}{llllllll} 
\tablecolumns{8} 
\tablewidth{0pc} 
\tablecaption{Table of Lenses} 
\tablehead{ \colhead{Lens}   & \colhead{$z_l$}    & \colhead{$z_Q$} & \colhead{Type}  & \colhead{$d_s$ }&\colhead{Images}& \colhead{$d_g$}&\colhead{Ref.}\\\colhead{}   & \colhead{}    & \colhead{} & \colhead{}  & \colhead{ ($\log d_s/''$)}&\colhead{}& \colhead{($''$)}&\colhead{}}
\startdata 
Q2237+030 &0.04 & 1.70 & Late\tablenotemark{a}& $-0.59\pm0.08$&(A,B,C,D) & (0.92, 0.97, 0.76, 0.88) & 1, 2\\
PG1115+080 &0.31 & 1.72 & Early& $-0.33\pm0.02$ & (A1,A2,B,C)&(1.18, 1.12, 0.95, 1.37) & 1, 3\\
B1422+231&  0.34&3.62 & Early& $-0.50\pm0.13$ & (A,B,C,D)& (0.95,0.89,1.04,0.36)& 1, 4\\
B1152+199 &0.44 & 1.02 & Late\tablenotemark{b}& $-0.8\pm0.2$\tablenotemark{c}& (A,B)&(1.14, 0.47) & 5, 6 \\
Q0142$-$100 & 0.49& 2.72& Early& $-0.29\pm0.02$& (A,B) & (1.86, 0.38)& 1, 7\\
B1030+071 & 0.60& 1.54& Early& $-0.35\pm0.06$ & (A,B)& (1.28, 0.58)& 1, 7\\
RXJ0911+0551 & 0.77& 2.80& Early& $-0.17\pm0.04$&(A,B,C,D) &(0.87, 0.97, 0.82, 2.24) & 8, 9\\
HE0512$-$3329&  0.93& 1.57& Late& $-1.0\pm0.3$\tablenotemark{c}& (A,B)& (0.035, 0.66)& 10, 11\\
MG0414+0534 & 0.96& 2.64& Early& $-0.11\pm0.08$ & (A1,A2,B,C) & (1.19, 1.17, 1.38, 0.96)& 1, 12\\
MG2016+112 &1.01 &  3.27& Early&$-0.66\pm0.05$ & (A,B)&(2.48, 1.25)& 1, 11 \\
\enddata 
\tablecomments{The table lists various properties of the lensing systems known from the literature.  The properties listed are the lens and quasar redshifts ($z_l$ and $z_Q$), the type and the scale length ($d_s$) of the lensing galaxy, and the image names and their distance from the center of the lens galaxy ($d_g$).  The scale length, $d_s$, is the effective radius of a de Vaucouleurs profile fit from \citet{rusin2003}.  The final column lists the references from which the values were obtained if they have not been previously quoted in the text.}

\tablenotetext{a}{The bulge is responsible for the lensing.}
\tablenotetext{b}{The spectra taken by \citet{myers} shows \ion{O}{2} emission line associated with the lens.  We therefore type B1152+199 as a late type galaxy. }
\tablenotetext{c}{For the purposes of the plotting of Figure~\ref{fig:A_scale} we adopt these scale lengths although they are not reported in the literature.}
\tablerefs{(1) \citet{rusin2003}, (2) \citet{rix}, (3) \citet{kristian}, (4) \citet{yee}, (5) \citet{myers}, (6) \citet{rusin}, (7) \citet{lehar}, (8) \citet{rusin_v1}, (9) \citet{burud}, (10) \citet{gregg}, (11) CASTLES (http://www.cfa.harvard.edu/castles/), (12) \citet{falco1997}.}
\label{tab:lenses}
\end{deluxetable*}

We create several sets of data consistent with the  \cardelli~with an effective microlensing parameter $0.01<s<0.3$ to test the effects of achromatic microlensing on our methods (see Fig. \ref{fig:sim_micro}).  We find that for very weak microlensing ($s=0.01$) there is no noticeable effect on the results (given $E(B-V) \ge 0.1$), in particular not when we include noise (of 0.05 mag) in our data points as the noise dominates the effects of the microlensing.  For a stronger microlensing signal ($0.05\lesssim s$) we find that the effects are indeed noticeable but to be able to quantify them it is crucial to be able to constrain the intrinsic magnitude ratio difference.  We find that a fit with a free intrinsic ratio can often give as good a fit (as measured by the $\chi^2_{\nu}$) as the fit which allows for correction due to the microlensing signal.  This is because the effects of the microlensing can in part be mimicked by shifting the whole data set up and down along the magnitude difference axis  by changing the intrinsic magnitude difference if the ratio of the continuum emission to the total emission is similar for the different bands (see eq. (\ref{eq:extmicro})).

\section{Observations}
\label{sec:data}

\begin{figure}
\epsscale{1.0}
\plotone{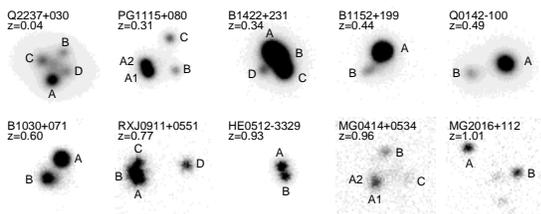}
\caption{Gallery of the 10 gravitational lensed quasars in the sample as they appear in the R-band VLT images.  The size of the stamps is $6\arcsec \times 6\arcsec$. \label{fig:lenses}}
\end{figure}

The lens systems were chosen to fulfill the criteria that they have an image separation larger than 0\farcs4 to ensure that the images of the quasar could be resolved, that they have declination $\delta<33\arcdeg$ to be visible with the ESO Very Large Telescope (VLT) at Paranal observatory and that the lens and quasar redshifts be known in order to reduce the number of unknowns when fitting the extinction curve.  At the time of the application, this left us with ten systems, five doubly imaged quasars (doubles) and five quadruply imaged quasars (quads).  The images are labeled according to the CfA-Arizona-Space-Telescope-LEns-Survey (CASTLES)\footnote{http://www.cfa.harvard.edu/castles/} notation (see Fig.~\ref{fig:lenses}).

Multi waveband imaging observations of the 10 gravitational lens systems was obtained with the VLT.   A list of the systems and their main properties known from the literature is given in Table~\ref{tab:lenses}, and a gallery of how they appear in the VLT observations is shown in Figure~\ref{fig:lenses}.  Optical observations (in the $U$, $B$, $V$, $R$, $I$ and $z$ band) was carried out with the FORS1 instrument (which with the high resolution collimator has a pixel scale of 0\farcs1), and near infrared (NIR) observations (in the $J_s$, $H$ and $K_s$ band) were carried out with the ISAAC instrument (which has a pixel scale of $0\farcs147$).

\begin{deluxetable*}{lrrrrrrrrrr} 
\tablecolumns{11} 
\tablewidth{0pc} 
\tablecaption{Overview of observations} 
\tablehead{ \colhead{Lens} & \colhead{$U$}   & \colhead{$B$}    & \colhead{$V$} & \colhead{$R$} & \colhead{$I$} & \colhead{$z$} & \colhead{$J_s$} & \colhead{$H$} & \colhead{$K_s$} & \colhead{Delay}\\ & \colhead{(s)}& \colhead{(s)}& \colhead{(s)}& \colhead{(s)}& \colhead{(s)}& \colhead{(s)}& \colhead{(s)}& \colhead{(s)}& \colhead{(s)} & \colhead{(d)}}
\startdata 
Q2237+030 &  180 & 60   & 60   & 40   & 60   &  40\tablenotemark{b} &  \nodata   & 60   & 60   & $-13$  \\ 
PG1115+080 & 30  & 15   & 9    & 9    & 9    &  9\tablenotemark{a}  & 180 & 300  & 180  & $-$20  \\ 
B1422+231 & 180  & 60   & 40   & 40   & 40   &  40\tablenotemark{b} &  \nodata   & 120  & 120  & 48  \\ 
B1152+199 & 3000 & 180  & 80   & 40   & 60   &  60\tablenotemark{b}  & 60  & 60   & 60   & 47  \\ 
Q0142$-$100 &  60  & 60   & 40   & 40   & 60   &  60\tablenotemark{b}&   \nodata  &   \nodata   &  \nodata & \nodata \\ 
B1030+071 & 9300 & 3600 & 3000 & 1400 & 1600 & 2000\tablenotemark{a} & 840 & 1280 & 2470 & $-$5   \\ 
RXJ0911+0551& 900& 30   & 15   & 15   & 25   &  15\tablenotemark{a}  & 180 & 216  & 360  & 5   \\ 
HE0512$-$3329 & 30 & 15   & 9    & 9    & 9    &  9\tablenotemark{a}  & 216 & 216  & 216  & $-$18  \\ 
MG0414+0534 &   \nodata & 5400 & 4200 & 270  & 60   &  120\tablenotemark{a}  & 240 & 240  & 240  & $-$11  \\ 
MG2016+112 & 2400& 1200 & 720  & 300  & 400  &  600\tablenotemark{b} & 480 & 960  & 1500 & $-$71  \\ 
\enddata 
\tablecomments{Total exposure time (in seconds) and delay between optical and NIR observations (in days), where a negative value denotes that the NIR were carried out before the optical.}
\tablenotetext{a}{The observing band was $z_{special}$.}
\tablenotetext{b}{The observing band was $z$ (Gunn)}
\label{obs}
\end{deluxetable*}

\begin{deluxetable*}{lllllllllll}
\tablecolumns{11}
\tablewidth{0pc}
\tablecaption{}
\tablehead{ \colhead{Lens} & \colhead{Image} & \colhead{$U$}   & \colhead{$B$}    & \colhead{$V$} & \colhead{$R$} & \colhead{$I$} & \colhead{$z$} & \colhead{$J_s$} & \colhead{$H$} & \colhead{$K_s$}}
\startdata
Q2237+030 & A &1.00    &     1.00    &     1.00   &     1.00   &      1.00     &    1.00     &    \nodata   &   1.00   &     1.00 \\
& $\sigma_A$& 0.02  &   0.02   &  0.01 &   0.01  &   0.02 &   0.02 &   \nodata  &   0.01 &  0.01\\
& B & 0.36 &       0.24  &     0.30    &    0.39    &   0.30   &    0.32  &      \nodata   &  0.33  &      0.36\\
&  $\sigma_B$& 0.02    & 0.02 &   0.01 &    0.01  &   0.02    &0.02  & \nodata    &0.01   & 0.01\\
& C & 0.26     &  0.35  &     0.40  &      0.29 &     0.37  &      0.45    &   \nodata  &   0.47  &     0.41\\
&  $\sigma_C$& 0.02  &  0.02 &  0.01  &  0.01  &  0.02 & 0.02&    \nodata   &  0.01   &0.01\\
& D &0.24   &     0.21  &     0.29   &     0.27     &   0.27        & 0.35     &  \nodata   &   0.38  &     0.35\\
&  $\sigma_D$&0.02  &0.02 &   0.01  &  0.01  &   0.02     & 0.02     &\nodata   &  0.01 &  0.01\\

PG1115+080 &A1& 1.00 &        1.000   &      1.000    &     1.000 &        1.000   &      1.00   &      1.000   &      1.000 &      1.000\\
&$\sigma_{A1}$&0.01   &0.006 &   0.006&   0.005  & 0.003   & 0.01 &  0.001  &  0.002 & 0.002\\
&A2&0.63 &       0.615   &     0.676   &     0.675&        0.710&       0.67 &       0.747 &       0.710&        0.771\\
&$\sigma_{A2}$&0.02  &   0.009 &   0.007  &0.008&   0.004  &  0.02  &  0.002 &   0.003 &  0.002\\
&B&0.26&        0.246&       0.253&      0.257&       0.262&      0.27     &   0.268  &     0.270  &     0.266\\
&$\sigma_{B}$&0.02&     0.009 &   0.008 &    0.008 &    0.004  &  0.02&     0.002  &   0.003   &  0.003\\
&C&0.18     &  0.17    &    0.17   &     0.17     &   0.170  &      0.17     &   0.177     &   0.178     & 0.176\\
&$\sigma_{C}$&0.02&    0.01 & 0.01 &   0.01 &    0.006 &    0.01 &    0.003  &  0.005   &   0.005\\

B1422+231  & A &0.105   &    0.15 &      0.16 &       0.15    &   0.18   &   0.16     &  \nodata  &  \nodata  &   0.31\tablenotemark{a}\\
&$\sigma_A$&0.008&  0.02&     0.01  &   0.01 &   0.02    & 0.01  &  \nodata   & \nodata  &   0.02\tablenotemark{a}\\
&B&0.09 &      0.14    &    0.16 &      0.15 &      0.17 &       0.16 &      \nodata &   \nodata&     0.35\tablenotemark{a}\\
&$\sigma_B$&0.01 &   0.02 & 0.02 &  0.02   & 0.02  &  0.01  &  \nodata   & \nodata&     0.02\tablenotemark{a}\\
&C&0.08     &   0.10 &      0.10   &    0.09  &    0.11  &     0.10   &    \nodata   & \nodata &    0.18\tablenotemark{a}\\
&$\sigma_C$&0.02  & 0.03 &    0.04 & 0.03  & 0.04   & 0.01  &  \nodata    &\nodata  &   0.02\tablenotemark{a}\\
&D&0.01    &    0.004  &      0.005   &     0.005   &     0.007 &       0.006 &      \nodata  &  \nodata    & 0.02\tablenotemark{a}\\
&$\sigma_D$&0.01&    0.018&     0.016&   0.014 &  0.019  & 0.013&   \nodata &   \nodata &    0.02\tablenotemark{a}\\

B1152+199 & A &\nodata  &    1.000 &       1.000   &      1.000 &       1.000    &     1.000  &      1.000     &    1.000  &       1.000\\
&$\sigma_A$&\nodata  &    0.007  &       0.002     &    0.002   &      0.003  &       0.005  &      0.004     &    0.002        & 0.004\\
&B&\nodata    &  0.0024    &  0.010   &   0.018    &   0.044  &     0.074 &      0.172  &      0.293   &     0.289\\
&$\sigma_B$&\nodata   &   0.0006   &    0.001  &      0.001    &    0.002  &      0.003   &     0.002 &      0.004       &  0.003 \\

Q0142$-$100  & A  &1.000&         1.00&        1.0000&        1.0000&         1.0000&        1.0000  &      \nodata &   \nodata &   \nodata\\
&$\sigma_A$&0.008  & 0.01&    0.0003&    0.0002&    0.0002    &  0.0004  & \nodata  & \nodata  &  \nodata\\
&B&0.13&  0.12   &     0.134  &     0.138  &     0.152&   0.159 &      \nodata   & \nodata    &\nodata\\
&$\sigma_B$&0.03   & 0.01&     0.001&     0.001  &   0.001 &    0.002   & \nodata  &  \nodata  &  \nodata\\

B1030+071\tablenotemark{\dag} & A &\nodata    & 1.00 & 1.00&  1.00& 1.00& 1.00& 1.00    &     1.00&         1.00\\
&$\sigma_A$&\nodata  &    0.01&    0.01&   0.01&     0.01&    0.02&     0.02&    0.03  &  0.02\\
&B&\nodata    & 0.16 &   0.23 &  0.30 &  0.41&   0.64&  0.378    &    0.29   &     0.28\\
&$\sigma_B$&\nodata    & 0.01&  0.01&    0.01& 0.01&  0.02&    0.008&    0.01  &   0.01 \\

RXJ0911+0551 & A &1.000&    1.00\tablenotemark{b}&         1.00       &  1.00   &      1.00      & \nodata    & 1.000      &   1.000         &1.000\\
&$\sigma_A$&0.008&    0.02\tablenotemark{b}  &  0.01  &   0.01  &   0.02  &  \nodata  &   0.008&   0.008& 0.006\\
&B&0.62&       0.94\tablenotemark{b}&        0.73&        0.72&       0.74    &   \nodata  &   0.923&  0.919       &0.970\\
&$\sigma_B$&0.01&   0.02\tablenotemark{b}  &  0.02  &  0.02  &   0.02  &  \nodata  &  0.009 & 0.008     &0.006\\
&C&0.33&      0.51\tablenotemark{b}&      0.37&      0.40&      0.41&   \nodata   &  0.49& 0.51     &  0.50\\
&$\sigma_C$&0.02  &  0.03\tablenotemark{b}&    0.04&    0.03&    0.04&  \nodata   &  0.02&   0.01     &0.01\\
&D&0.37&       0.40\tablenotemark{b} &       0.38&        0.37&      0.37&    \nodata  &   0.39 &  0.44       & 0.40\\
&$\sigma_D$&0.01&    0.02\tablenotemark{b} &    0.02 &     0.01 &  0.02 &  \nodata   &  0.02& 0.02 &0.01 \\

HE0512$-$3329 & A & 1.000& 1.000&  1.000      &   1.0000    &    1.000 &  \nodata    & 1.0000  & 1.0000 &1.0000\\
&$\sigma_A$&0.008&    0.002&     0.002  &   0.0009   &  0.001  &  \nodata  &   0.0005  & 0.0004     &0.0005\\
&B&1.141&       1.021&         0.887     &   0.738     &  0.651       &\nodata    &  0.5813     & 0.5667  &      0.5593\\
&$\sigma_B$&0.007   &   0.002&   0.002&    0.001&     0.002  &  \nodata   & 0.0008& 0.0007     &0.0009\\

MG0414+0534 & A1 &\nodata    &\nodata   &  1.000&         1.000&        1.00    &     1.00\tablenotemark{c}     &    1.000       &  1.000      &   1.000 \\
&$\sigma_{A1}$&\nodata    & \nodata  &   0.005&   0.008 &  0.01&     0.02\tablenotemark{c} &   0.004&   0.003  &   0.001\\
&A2&\nodata  &  \nodata   &  0.41&      0.34&       0.39&      0.89\tablenotemark{c}     &    0.567     &   0.748     &   0.780\\
&$\sigma_{A2}$&\nodata  &  \nodata   &  0.01&    0.02&   0.03&     0.02\tablenotemark{c}&   0.008&   0.004&     0.003\\
&B&\nodata    &\nodata   &  0.818&      0.592&        0.50&       0.78\tablenotemark{c}&     0.416   &     0.394       & 0.385\\
&$\sigma_B$&\nodata  &  \nodata   &  0.005&    0.009&    0.01&     0.02\tablenotemark{c}&    0.006&  0.004    & 0.006\\
&C&\nodata    &\nodata  &   0.396   &    0.29&        0.24   &     0.34\tablenotemark{c}    &    0.18       & 0.179        &0.175\\
&$\sigma_C$&\nodata  &  \nodata  &   0.008   &  0.01&  0.02&    0.03\tablenotemark{c}&    0.01     &0.006   &  0.005 \\

MG2016+112 & A &\nodata   &  1.00& 1.00& 1.00&     1.00&     1.00\tablenotemark{d}& 1.0&  1.00& \nodata\\
&$\sigma_A$&\nodata  &   0.01&   0.01&    0.01&     0.01&      0.02\tablenotemark{d}&    0.1&  0.03   & \nodata\\
&B&\nodata     &0.60&      0.68&       0.81&      0.87     &   1.04\tablenotemark{d}       &  1.08 & 0.94     & \nodata\\
&$\sigma_B$&\nodata    & 0.02&    0.01&    0.01&     0.01   &  0.02\tablenotemark{d}  &   0.09 & 0.03 &  \nodata\\
\enddata
\tablecomments{Data table with the results from the deconvolution.  Missing data are either due to lack of observations (see overview in Table~\ref{obs}) or failure of the deconvolution to converge.}
\tablenotetext{\dag}{The photometry of the B component is believed to be contaminated by the lensing galaxy.}
\tablenotetext{a}{The images are not well resolved.  This system shows very weak extinction, and no detailed extinction curve analysis is performed, so the exclusion of this point should not affect our results.}
\tablenotetext{b}{Seeing too high to separate components (0\farcs70).}
\tablenotetext{c}{Very faint detection.}
\tablenotetext{d}{Very faint detection.}

\label{tab:original_data}
\end{deluxetable*}

\begin{figure}
\epsscale{1.0}
\plotone{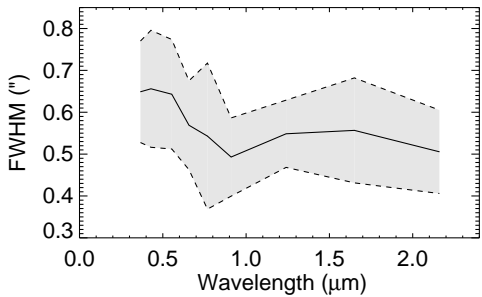}
\caption{Mean FWHM  measured in the VLT observations of the 10 systems, as a function of wavelength (full curve). The shaded area between the dashed curves indicates the RMS scatter around the mean. Approximately $80\%$ of the observations were carried out in $\lesssim 0.65 \arcsec$ seeing.  \label{fig:seeing}}
\end{figure}

The data were collected in excellent seeing conditions (FWHM~$\lesssim$~0\farcs65, see Figure~\ref{fig:seeing}) with mean seeing of $0\farcs57$ for the full data set.
Photometric conditions were not necessary since we are considering only relative photometry. 
An effort was made to carry out the different waveband observations of each system as close in time as possible to each other, to minimize the effects of time dependent intrinsic quasar variation and achromatic microlensing. 
For each system, the optical wavebands were observed on the same night in immediate succession, while the NIR observations were observed as close in time as scheduling allowed (the mean delay was 18 days, see Table~\ref{obs}). 
The effects of time delayed intrinsic variations between the individual images is thus reduced to a possible shift between the optical and NIR fluxes.
Details of the observations are summarized in Table~\ref{obs}.


\subsection{Data Reduction}
The individual optical data frames were bias subtracted and flat fielded using the standard ESO pipeline, and cleaned for cosmic rays using Laplacian edge detection \citep{vandokkum}.  In some of the frames the background was found to have large-scale gradients over the field. To properly account for this, we ran all the frames through the {\sc sextractor} software \citep{bertin}, with options set to save a full resolution interpolated background map, which was then subtracted from the science frames. 

The NIR data were reduced using a combination of the {\sc eclipse} software \citep{devillard} and the {\sc IRAF} ``Experimental Deep Infrared Mosaicing Software'' {\sc xdimsum}.  The {\sc eclipse} software was used to remove the effects of electrical ghosts from science and calibration frames, and to construct flat fields and bad-pixel maps from a series of twilight sky flats.  Sky subtraction and combination of the individual science frames was carried out with {\sc xdimsum}. 

\subsection{Deconvolution and photometry}
All photometry was carried out using the MCS deconvolution \citep{magain}.  This method uses a model PSF, measured directly from the data, to deconvolve the images, assuming that the gravitational lens systems can be decomposed into a number of point source (the quasar images) and a diffuse extended component (the lensing galaxy).  Positions and amplitudes of the quasar images are left free in the fit, and relative photometry can be derived from the best fitting amplitude.  No functional form was assumed for the diffuse component, which is a purely numerical component.  In the cases when there was more than one image of a given system available (in a given waveband), we performed simultaneous deconvolution of the individual images rather than deconvolving the combined image.

Photometric errors were estimated by the MCS algorithm, and  include photon noise and errors associated with deconvolution \citep{magain, burudb}.  A full list of the results from the MCS deconvolution is given in Table~\ref{tab:original_data}.  In addition to the MCS errors, we applied a 0.05 mag error on all the calculated magnitude differences to account for other sources of systematic noise.

We exclude data points from our sample where the deconvolution did not converge or where only one component was detected (7 data points).  In addition, we exclude 4 data points from further analysis.  These points are marked in Table~\ref{tab:original_data} and an explanation of the exclusion of each point is given in a footnote.  We also exclude from further analysis all the data taken for B1030+071 as the B component is heavily contaminated by the lens galaxy (separated by $0\farcs11\pm0\farcs01$ \citep{xan1}).  

\section{Results and Discussion}
\label{sec:results}

We open this chapter by presenting in \S~\ref{sec:res_ind} the results of our extinction curve analysis for each of the 10 lensing systems. We present the detailed analysis of systems where at least one image pair has a two sigma detection of extinction for one of the three applied extinction laws.  The systems are presented in order of increasing redshift.  We then move on to discussing in \S~\ref{sec:res_full} the overall results of our analysis, as well as statistical properties of the full sample.

\subsection{The individual systems}
\label{sec:res_ind}

\subsubsection{Q2237+030}
Q2237+030 was discovered by \citet{huchra} and consists of a quasar at redshift $z=1.70$ and a spiral lensing galaxy at $z=0.04$ making it the nearest known lensing galaxy to date.  The system was later resolved into four images forming an Einstein cross with the lensing galaxy in the middle  \citep{yee88, schneider88}.  \citet{schneider88} modeled the system in detail and found a predicted time delay of order of one day between the images and amplifications of $4.6,4.5,3.8$ and $3.6$ for images A, B, C and D, respectively.  \citet{falco96} studied the system with the VLA at radio wavelengths and obtained flux density ratios of $1.00,1.08,0.55$ and $0.77$ for images A, B, C, D, respectively.  Q2237+030 has previously been noted in the literature as having high variability which is uncorrelated between the four images \citep[see e.g.,][]{irwin, corrigan}.  

\begin{figure}
\epsscale{1.0}
\plotone{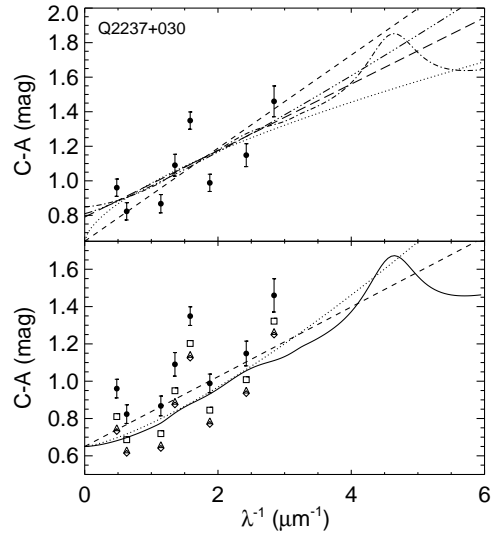}
\caption{Q2237+030, C$-$A:  The upper panel shows the data points and the best fit extinction curves.  The lower panel shows the original data points and their shift due to a microlensing signal, along with the best fit through the shifted points.  The parameters of the fits are given in Table~\ref{tab:q2237ac}.  \textit{Annotation:}  Filled circles are the original data points with error bars.  The curves correspond to the \cardelli~(eq. \ref{eq:car}) with $\Delta\hat{m}$ free (dash-dot) and fixed (solid), the power law (eq. \ref{eq:alpha}) with  $\Delta\hat{m}$ free (dash-dot-dot-dot) and fixed (dotted) and the linear law (eq. \ref{eq:lambda}) with $\Delta\hat{m}$ free  (long dash) and fixed (short dash).  Shifted data points due to a microlensing signal are plotted in open boxes (\cardelli), triangles (power law) and diamonds (linear law).\label{fig:q2237c}}
\end{figure}

\begin{figure}
\epsscale{1.0}
\plotone{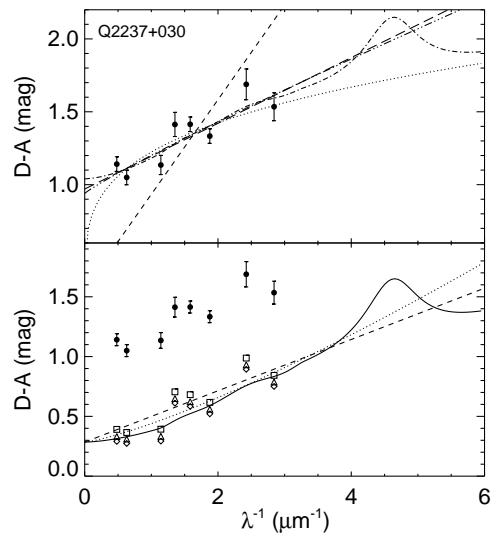}
\caption{Q2237+030, D$-$A:  The upper panel shows the data points and the best fit extinction curves.  The lower panel shows the original data points and their shift due to a microlensing signal, along with the best fit through the shifted points.  The parameters of the fits are given in Table~\ref{tab:q2237ad}.  See the caption of Figure \ref{fig:q2237c} for annotation overview. \label{fig:q2237d}}
\end{figure}

This system is difficult to interpret as it shows a lot of scatter in the points which cannot be explained by extinction alone nor microlensing (see Figures~\ref{fig:q2237c} and \ref{fig:q2237d}).  None of the extinction laws we apply give a good fit to the data for any pair of images (as the redshift of this system is very low, $z=0.04$, we do not expect to see much difference between the quality of the fits for the different extinction laws, see \S~\ref{sec:pureext}).  All the image pairs, except C$-$A and D$-$A, yield $A(V)$ consistent with zero.  The data points and fits for C$-$A and D$-$A can be seen in Figures~\ref{fig:q2237c} and \ref{fig:q2237d} and the parameters of the fits in Tables~\ref{tab:q2237ac} and \ref{tab:q2237ad}.

\begin{deluxetable*}{lrrrrrrr} 
\tablecolumns{7} 
\tablewidth{0pc} 
\tablecaption{Extinction curve fit results for Q2237+030: C$-$A} 
\tablehead{ \colhead{Extinction} & \colhead{$A(V)$}   & \colhead{$R_V$}    & \colhead{$\alpha$} & \colhead{$\Delta\hat{m}$} & \colhead{$s$} & \colhead{$\chi^2_{\nu}$}}
\startdata 
\cardellitab & $ 0.53\pm0.03$ & $>7$  &\nodata  & \textit{0.65}& \nodata & $3.6\pm0.4$\\  
\cardellitab & $0.29\pm0.06$ & $2.9\pm1.4$ & \nodata & $0.85\pm0.04$ & \nodata & $3.4\pm0.4$\\ 
\cardellitab & $0.27\pm0.07$ & $2.6\pm 1.5$ & \nodata & \textit{0.65} & $-0.23\pm0.05$ & $3.4\pm0.4$\\ 
Power law  & $0.49\pm0.02 $ & \nodata & $0.6\pm0.1$ & \textit{0.65} & \nodata & $3.2\pm0.4$\\
Power law  & $0.3\pm0.5$   & \nodata & $1\pm4$ & $0.8\pm0.5$ & \nodata & $3.4\pm0.4$\\  
Power law  & $0.3\pm0.2$ & \nodata & $1\pm3$ & \textit{0.65} & $-0.2\pm0.1$ & $3.4\pm0.4$\\  
Linear law& $0.49\pm0.02$ & \nodata & \textit{1.0} & \textit{0.65} & \nodata & $3.2\pm0.4$\\  
Linear law& $0.35\pm0.05$ & \nodata & \textit{1.0} & $0.78\pm0.04$ &\nodata & $3.1\pm0.4$\\  
Linear law& $0.34\pm0.05$ & \nodata & \textit{1.0} & \textit{0.65} & $-0.15\pm0.05$ & $3.1\pm0.4$\\
\enddata 
\tablecomments{ The extinction of image C compared to image A.  Numbers quoted in italics were fixed in the fitting procedure.}
\label{tab:q2237ac}
\end{deluxetable*}

\begin{deluxetable*}{lrrrrrrr} 
\tablecolumns{7} 
\tablewidth{0pc} 
\tablecaption{Extinction curve fit results for Q2237+030:  D$-$A} 
\tablehead{ \colhead{Extinction} & \colhead{$A(V)$}   & \colhead{$R_V$}    & \colhead{$\alpha$} & \colhead{$\Delta\hat{m}$} & \colhead{$s$} & \colhead{$\chi^2_{\nu}$}}
\startdata 
\cardellitab & $ 1.23\pm0.03$ & $>7$  &\nodata  & \textit{0.28}& \nodata & $7.1\pm0.4$\\  
\cardellitab & $0.35\pm0.06$ & $3.1\pm1.5$ & \nodata & $1.05\pm0.05$ & \nodata & $1.9\pm0.4$\\ 
\cardellitab & $0.28\pm0.07$ & $2.1\pm 1.2$ & \nodata & \textit{0.28} & $-0.85\pm0.05$ & $1.9\pm0.4$\\ 
Power law  & $1.11\pm0.03 $ & \nodata & $0.27\pm0.04$ & \textit{0.28} & \nodata & $1.9\pm0.4$\\
Power law  & $0.4\pm1.1$   & \nodata & $0.9\pm0.4$ & $0.9\pm1.1$ & \nodata & $2.0\pm0.4$\\  
Power law  & $0.3\pm0.4$ & \nodata & $1.3\pm0.6$ & \textit{0.28} & $-0.8\pm0.4$ & $2.0\pm0.4$\\  
Linear law& $1.18\pm0.03$ & \nodata & \textit{1.0} & \textit{0.28} & \nodata & $5.9\pm0.4$\\  
Linear law& $0.41\pm0.06$ & \nodata & \textit{1.0} & $0.97\pm0.05$ &\nodata & $1.9\pm0.4$\\  
Linear law& $0.39\pm0.06$ & \nodata & \textit{1.0} & \textit{0.28} & $-0.75\pm0.05$ & $1.9\pm0.4$\\
\enddata 
\tablecomments{  The extinction of image D compared to image A.  Numbers quoted in italics were fixed in the fitting procedure.}
\label{tab:q2237ad}
\end{deluxetable*}

We use the radio measurements of \citet{falco96} to fix $\Delta\hat{m}$ and in particular for the D$-$A image pair this changes the results significantly.  As the flux density ratios in the radio agree with the model predictions for the D$-$A image pair it is interesting to note that none of the extinction laws we apply give good fits to the data unless we allow for corrections due to achromatic microlensing (see Figures~\ref{fig:q2237c} and \ref{fig:q2237d} and Tables~\ref{tab:q2237ac} and \ref{tab:q2237ad}).  This might suggest that either D is demagnified by a strong microlensing signal, or that image A is magnified (or both) and the residual `extinction curve' which we are fitting may be effects of chromatic microlensing.  The same effect, but not as strong, is seen in the C$-$A image, again suggesting a slight demagnification of C, a magnification of A or a combination of the two.  In previous microlensing studies, component D has not been seen to have as strong a microlensing signal as A and C have \citep{irwin, alcalde, gil-merino} so therefore it is perhaps more likely that we are seeing the magnification of image A.

Another explanation for the shift in $\Delta\hat{m}$ might be intrinsic variations of the quasar components as discussed in \S \ref{sec:lensing}.  Such variability could introduce an overall shift of the data, resulting in inaccurate estimates of $\Delta\hat{m}$.

By inspecting the $V$-band lightcurves from \citet{kochanek2004} we see that component A is indeed in a bright phase at the time of our observations (Julian date of around 2451670).  Component D is fairly stable but component C is getting dimmer climbing down from a peak in its brightness and is still fairly bright, perhaps making the C$-$A shift in $\Delta\hat{m}$ less prominent than the D$-$A shift.

\subsubsection{PG1115+080}
PG1115+080 is a  multiply imaged system discovered by \citet{weynman} as a triply imaged system with the quasar at redshift $z=1.72$.  The A component was later resolved into two separate images, A1 and A2, by \citet{hege} making the system a quad.  The lensing galaxy was located by \citet{christian} and is an early type galaxy \citep{rusin}.  Its redshift and that of three neighboring galaxies were determined to be at $z=0.31$ by \citet{kundica}.  The time delays between the components were determined by \citet{schechter}.

\begin{deluxetable*}{lrrrrr} 
\tablecolumns{6} 
\tablewidth{0pc} 
\tablecaption{Extinction curve fit results for PG1115+080} 
\tablehead{ \colhead{Extinction} & \colhead{$A(V)$}   & \colhead{$R_V$}    & \colhead{$\alpha$} & \colhead{$\Delta\hat{m}$}  & \colhead{$\chi^2_{\nu}$}}
\startdata 
\cardellitab & $0.12\pm0.06$ & $3.3\pm1.9$ & \nodata & $0.29\pm0.05$  & $1.2\pm0.3$\\ 
Power law  & $0.3\pm2.0$   & \nodata & $0.5\pm1.5$ & $0.1\pm2.0$ & $1.2\pm0.3$\\  
Linear law& $0.13\pm0.03$ & \nodata & \textit{1.0} & $0.28\pm0.04$  & $1.2\pm0.3$\\  
\enddata 
\tablecomments{  The extinction of image A2 compared to reference image A1.  Numbers quoted in italics were fixed in the fitting procedure.}
\label{tab:pg1115ab}
\end{deluxetable*}

\begin{figure}
\epsscale{1.0}
\plotone{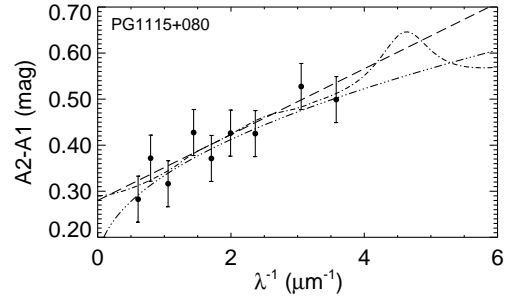}
\caption{PG1115+080, A2$-$A1:  Best fit extinction curves for A2$-$A1.  The parameters of the fits can be seen in Table~\ref{tab:pg1115ab}.  See the caption of Figure \ref{fig:q2237c} for annotation overview.\label{fig:pg1115}}
\end{figure}

The system shows very weak differential extinction for all the images (all but A2$-$A1 have $A(V)$ equal to $0$ within two sigma, see Table~\ref{tab:extinction}).  The data points and fits for A2$-$A1 are shown on Figure  \ref{fig:pg1115} and the parameters of the fits in Table  \ref{tab:pg1115ab}.  The low extinction signal is in agreement with the results of \citet{falco1999}.

\subsubsection{B1422+231}
B1422+231 is a quadruply imaged system first discovered by \citet{patnaik} in the JVAS survey and confirmed to be a lensing system by \citet{lawrence92}.  The lensing system consists of an early type main galaxy \citep{yee} at $z=0.34$ \citep{kundic, tonry98} and five nearby galaxies \citep{remy, bechtold}.  The quasar is at a redshift of $z=3.62$ and the maximum image separation is  $1.3''$  \citep{patnaik}.  The images show intrinsic variability which has been used to determine the time delay by studying radio light curves \citep{patnar}.

We only use three of the four images in our analysis as D was too faint to give usable results (all the visible bands gave zero detection).  As for the other components they show very weak differential extinction and give very weak constraints on the differential extinction curves.  All the fits have $A(V)$ consistent with zero (see Table~\ref{tab:extinction}).  We also get $A(V)$ consistent with zero when we fix $\Delta\hat{m}$ (where the values for the $\Delta\hat{m}$ are taken to be the average between those deduced by \citet{patnaik} in the $5~GHz$ and $8~GHz$ bands).  The low differential extinction between the images is in agreement with the results of \citet{falco1999}.

\subsubsection{B1152+199}
\begin{deluxetable*}{lrrrrrrr} 
\tablecolumns{7} 
\tablewidth{0pc} 
\tablecaption{Extinction curve fit results for B1152+199} 
\tablehead{ \colhead{Extinction} & \colhead{$A(V)$}   & \colhead{$R_V$}    & \colhead{$\alpha$} & \colhead{$\Delta\hat{m}$} & \colhead{$s$} & \colhead{$\chi^2_{\nu}$}}
\startdata 
\cardellitab & $ 2.03\pm0.03$ & $1.61\pm0.05$ &\nodata  & \textit{1.18}& \nodata & $2.7\pm0.4$\\  
\cardellitab & $2.43\pm0.09$ & $2.1\pm0.1$ & \nodata & $0.85\pm0.07$ & \nodata & $2.0\pm0.4$\\ 
\cardellitab & $2.41\pm0.09$ & $2.0\pm 0.1$ & \nodata & \textit{1.18} & $0.32\pm0.07$ & $2.1\pm0.4$\\ 
Power law  & $2.01\pm0.03 $ & \nodata & $1.98\pm0.04$ & \textit{1.18} & \nodata & $4.0\pm0.4$\\
Power law  & $2.7\pm0.1$   & \nodata & $1.45\pm0.08$ & $0.6\pm0.1$ & \nodata & $2.8\pm0.4$\\  
Power law  & $2.6\pm0.1$ & \nodata & $1.52\pm0.07$ & \textit{1.18} & $0.6\pm0.1$ & $3.2\pm0.4$\\  
Linear law& $1.94\pm0.03$ & \nodata & \textit{1.0} & \textit{1.18} & \nodata & $10.0\pm0.4$\\  
Linear law& $3.57\pm0.07$ & \nodata & \textit{1.0} & $-0.23\pm0.06$ &\nodata & $3.4\pm0.4$\\  
\enddata 
\tablecomments{  The extinction of image B compared to reference image A.  Numbers quoted in italics were fixed in the fitting procedure.}
\label{tab:b1152}
\end{deluxetable*} 

B1152+199 is a doubly imaged system first discovered by \citet{myers} in the CLASS survey with a background quasar at $z=1.02$, a lensing galaxy at $z=0.44$ and image separation of $1\farcs56$.  It was observed in radio wavelengths (at frequencies  $1.4,5,8.4$ and $15$~$GHz$) by \citet{rusin}.  The extinction curve has previously been studied and fitted by a \cardelli~with $1.3\leq R_V\leq2.0$   and $E(B-V)\sim1$ \citep{toft} suggesting that it is a heavily extinguished system.

\begin{figure}
\epsscale{1.0}
\plotone{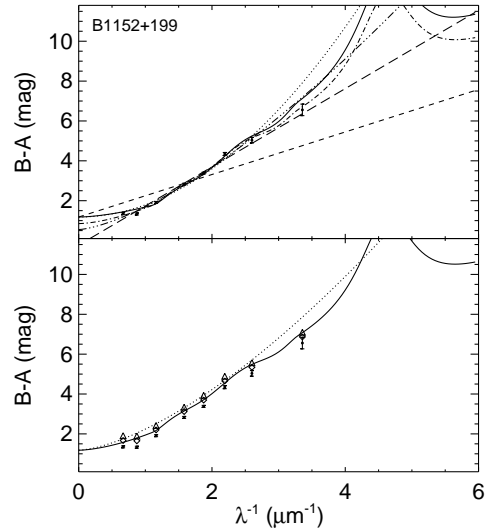}
\caption{B1152+199:  The upper panel shows the data points and the best fit extinction curves to them.  The lower panel shows the original data points and their shift due to a microlensing signal.  The parameters of the fits are given in Table~\ref{tab:b1152}.  See the caption of Figure \ref{fig:q2237c} for annotation overview. \label{fig:b1152}}
\end{figure}

B1152+199 shows a very strong extinction signal as can be seen in Figure~\ref{fig:b1152} and Table~\ref{tab:b1152}.  It has the strongest extinction signal of all ten systems with $A(V)=2.43\pm0.09,2.7\pm0.1,3.57\pm0.07$ at $\chi^2_\nu=2.0,2.8,3.4$ for the \cardelli, power law and linear law respectively.  Using the radio measurements of \citet{rusin} to fix $\Delta\hat{m}$ we similarly get $A(V)=2.03\pm0.03,2.01\pm0.03,1.94\pm0.03$ at $\chi^2_\nu=2.7,4.0,10.0$.  We also analyze the data with respect to a possible achromatic microlensing signal, keeping $\Delta\hat{m}$ fixed.  This yields a non-zero microlensing correction for the \cardelli~and power law of $s=0.32\pm0.07,0.6\pm0.1$ and $A(V)=2.41\pm0.09,2.6\pm0.1$ at $\chi^2_\nu=2.1,3.2$ respectively.  The best fit for the linear law lies outside the validity of the method with $s>1$ (see \S~\ref{sec:micro}) which would correspond to a microlensing signal of $>1$ mag.

It is clear that in all cases the \cardelli~provides the best fit to the data suggesting Galactic type dust although the best fit $R_V$ values are lower than those commonly seen in the Milky Way.  It is possible that the measured $R_V$ value is being lowered by a non-zero extinction in the A image provided it has a higher value of $R_V$ (see discussion in \S~\ref{sec:extboth}).  However, given the very strong extinction signal this would require very strong extinction along both lines of sight in addition to a strong differential signal.  This is unlikely given the fact that component A is at more than twice the distance from the center of the lensing galaxy than component B with A at $1\farcs14$ and B at $0\farcs47$ from the center \citep{rusin}. Measurements in the X-ray further suggest that the A component is non-extinguished (K. Pedersen et al., 2006, in preparation).

\subsubsection{Q0142$-$100}
\label{sec:q0142}
Q0142$-$100 is a doubly imaged system first discovered by \citet{surdej} and also known in the literature as UM 673.  The quasar is at a redshift of $z=2.72$ \citep{macalpine} and the lensing galaxy, which is of early type \citep{rusin}, is at a redshift of $z=0.49$ \citep{surdej}.   \citet{wisotzki} studied this system using spectrophotometric observations and found signs of differential extinction but no microlensing.

\begin{figure}
\epsscale{1.0}
\plotone{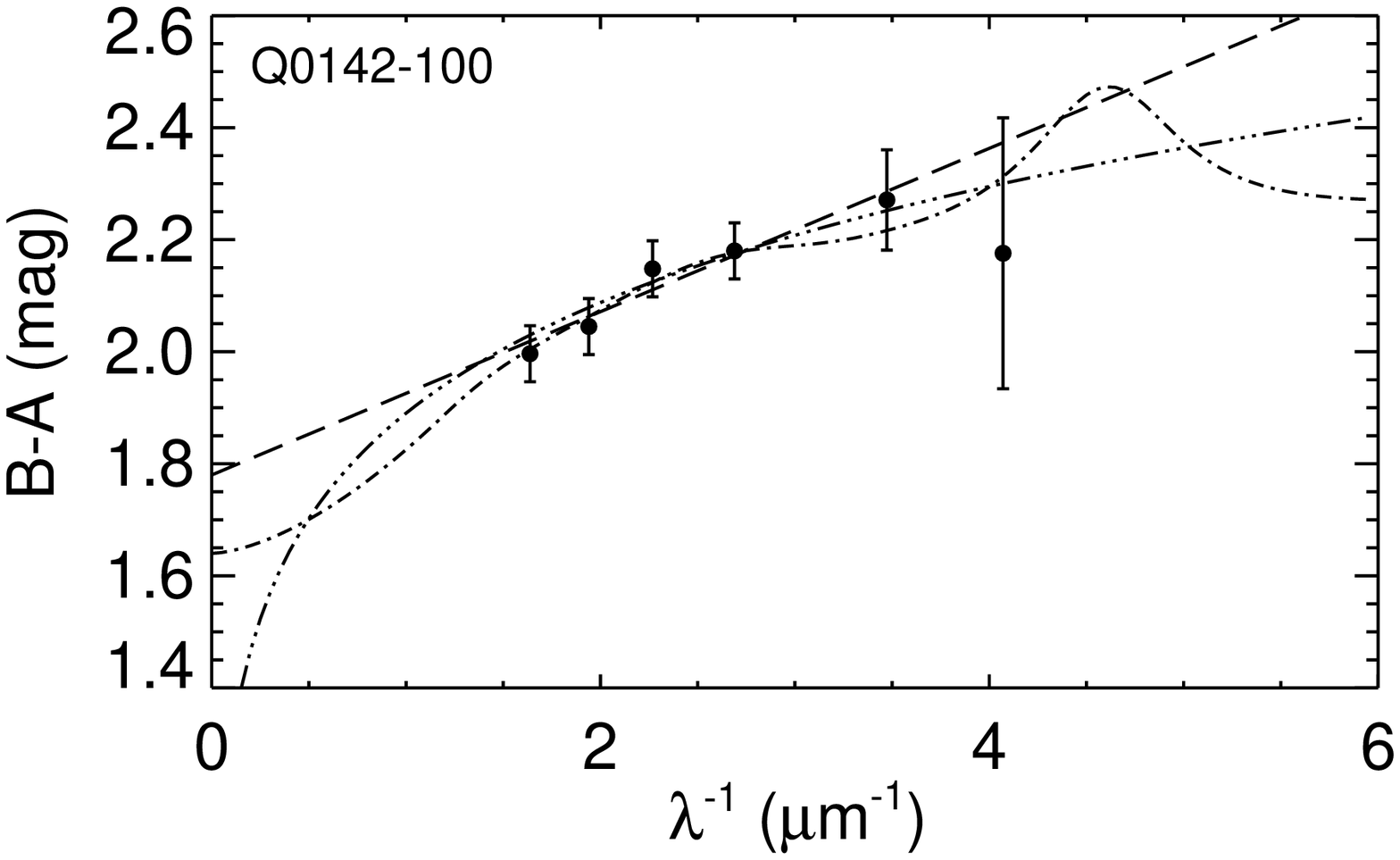}
\caption{Q0142$-$100:  The plot shows the data points and the best fit to them.  The parameters of the fits can be seen in Table~\ref{tab:q0142}.  See the caption of Figure \ref{fig:q2237c} for annotation overview. \label{fig:q0142}}
\end{figure}

The data points and our fits for Q0142$-$100 can be seen in Figure~\ref{fig:q0142} and the parameters of the fits in Table~\ref{tab:q0142}.  All the fits give similar $\chi^2_\nu$ but the parameters, in particular for the power law, are poorly constrained due to the lack of data points (we did not get any measurements in the infrared for this system).  The extinction is high for an early type galaxy and is not consistent with that found by \citet{falco1999} who found negligible extinction.  We suspect that the data are being contaminated by the lens galaxy as the B component is located near the galaxy center (at $0\farcs38$) and the seeing was not optimal for this system (the mean seeing was $0\farcs87$ compared to $0\farcs57$ for the full data set). As there are no published radio measurements available we do not have constraints on $\Delta\hat{m}$ to analyze the system with respect to a possible microlensing signal.  

\begin{deluxetable*}{lrrrrr} 
\tablecolumns{6} 
\tablewidth{0pc} 
\tablecaption{Extinction curve fit results for Q0142$-$100} 
\tablehead{ \colhead{Extinction} & \colhead{$A(V)$}   & \colhead{$R_V$}    & \colhead{$\alpha$} & \colhead{$\Delta\hat{m}$} &\colhead{$\chi^2_{\nu}$}}
\startdata 
\cardellitab & $0.40\pm0.03$ & $4.8\pm0.7$ & \nodata & $1.64\pm0.04$ & $1.2\pm0.5$\\ 
Power law  & $4.1\pm3.8$   & \nodata & $0.1\pm0.8$ & $-2.1\pm3.8$ & $1.1\pm0.5$\\  
Linear law& $0.27\pm0.08$ & \nodata & \textit{1.0} & $1.8\pm0.1$  & $1.1\pm0.4$\\  
\enddata 
\tablecomments{  The extinction of image B compared to reference image A.  Numbers quoted in italics were fixed in the fitting procedure.}
\label{tab:q0142}
\end{deluxetable*} 

\subsubsection{B1030+071}
B1030+071 is a doubly imaged system first discovered by \citet{xan1} in the JVAS survey.  They monitored the system in radio wavelengths, finding that the flux density ratios between image A and B range from 12.0 to 18.8 and seem to vary with both time and frequency.  The redshift of the background source was determined to be at $z=1.54$ and the redshift of the lensing object to be at $z=0.60$ \citep{fassnacht}.  \citet{falco1999} determined a differential extinction of $E(B-V)=0.02\pm0.04$ assuming a fixed $R_V=3.1$ \cardelli.

We were unable to perform an extinction analysis on this system as the deconvolution did not succeed in separating the B component from the main lens galaxy (separated by $0\farcs11\pm0\farcs01$ \citep{xan1}) making the photometric values unreliable.  For a further study of the extinction of this system higher resolution images would be required.

\subsubsection{RXJ0911+0551}

\begin{deluxetable*}{lrrrrrrr} 
\tablecolumns{7} 
\tablewidth{0pc} 
\tablecaption{Extinction curve fit results for RXJ0911+0551: B} 
\tablehead{ \colhead{Image pair} & \colhead{Extinction} & \colhead{$A(V)$}   & \colhead{$R_V$}    & \colhead{$\alpha$} & \colhead{$\Delta\hat{m}$}  & \colhead{$\chi^2_{\nu}$}}
\startdata 
B$-$A & \cardellitab & $ 0.33\pm0.06$ & $4.9\pm0.6$ &\nodata  & $-0.10\pm0.05$ &  $1.3\pm0.4$\\ 
B$-$D  & \cardellitab & $ 0.20\pm0.08$ & $4.1\pm1.1$ &\nodata  & $-1.00\pm0.06$ & $1.6\pm0.5$\\
B$-$A & Power law & $0.9\pm3.7$ & \nodata & $0.3\pm0.3$ & $-0.7\pm3.7$ & $1.3\pm0.4$ \\
B$-$D & Power law & $0.2\pm1.4$ & \nodata & $0.8\pm0.6$ & $-1.1\pm1.4$ & $1.6\pm0.5 $\\
B$-$A & Linear law & $0.23\pm0.03$ & \nodata & \textit{1.0} & $-0.04\pm0.04$ & $1.5\pm0.4$\\
B$-$D & Linear law & $0.17\pm0.03$ & \nodata & \textit{1.0} & $-1.00\pm0.04$ & $1.5\pm0.4$
\enddata
\tablecomments{  The extinction of image B compared to reference images A and D.  Numbers quoted in italics were fixed in the fitting procedure.} 
\label{tab:rxj0911b}
\end{deluxetable*}

\begin{deluxetable*}{lrrrrrrr} 
\tablecolumns{7} 
\tablewidth{0pc} 
\tablecaption{Extinction curve fit results for RXJ0911+0551: C} 
\tablehead{ \colhead{Image pair} & \colhead{Extinction} & \colhead{$A(V)$}   & \colhead{$R_V$}    & \colhead{$\alpha$} & \colhead{$\Delta\hat{m}$}  & \colhead{$\chi^2_{\nu}$}}
\startdata 
C$-$A & \cardellitab & $ 0.20\pm0.10$ & $3.3\pm1.5$ &\nodata  & $0.66\pm0.07$ &  $1.2\pm0.4$\\ 
C$-$D  & \cardellitab & $ 0.09\pm0.08$ & $2.2\pm1.5$ &\nodata  & $0.26\pm0.06$ & $1.3\pm0.4$\\
C$-$A & Power law & $0.3\pm2.1$ & \nodata & $0.9\pm0.5$ & $0.6\pm2.1$ & $1.2\pm0.4$ \\
C$-$D & Power law & $0.1\pm1.8$ & \nodata & $1.3\pm1.5$ & $03\pm1.8$ & $1.2\pm0.4$\\
C$-$A & Linear law & $0.23\pm0.04$ & \nodata & \textit{1.0} & $0.62\pm0.04$ & $1.1\pm0.4$\\
C$-$D & Linear law & $0.17\pm0.04$ & \nodata & \textit{1.0} & $0.31\pm0.05$ & $1.2\pm0.4$
\enddata 
\tablecomments{  The extinction of image C compared to reference images A and D.  Numbers quoted in italics were fixed in the fitting procedure.}
\label{tab:rxj0911c}
\end{deluxetable*}

RXJ0911+0551 is a  multiply imaged system first discovered by \citet{bade} in the ROSAT All-Sky Survey with the quasar at $z=2.80$.  It was later studied by \citet{burud} who resolved the system into four images and found that large external shear, possibly due to a cluster, was required to explain the image configuration.  \citet{kneib} confirmed that the lensing galaxy belongs to a cluster at $z=0.769$.  Observed reddening in at least two (images B and C) of the four images suggest differential extinction by the early type lensing galaxy \citep{burud}.   \citet{hjorth} measured the time delay of the system between images A,B,C on the one hand  and D on the other and found the time delay to be $146\pm8$~days ($2\sigma$).

\begin{figure}
\epsscale{1.0}
\plotone{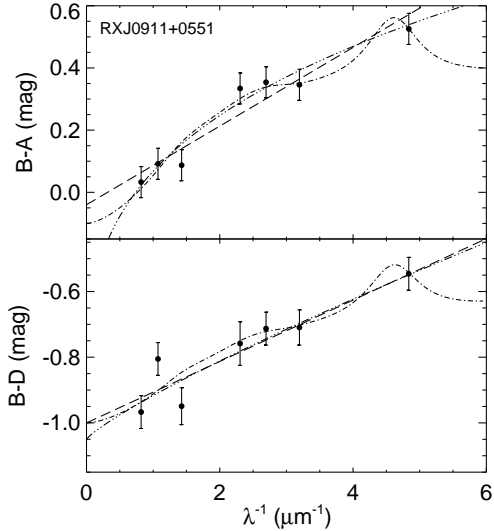}
\caption{RXJ0911+0551, B:  The upper panel shows the data points and the fits for the extinction of image B compared to image A.  The lower panel shows the corresponding plot for image B compared to image D.  The parameters of the fits can be seen in Table~\ref{tab:rxj0911b}.  See the caption of Figure \ref{fig:q2237c} for annotation overview.    \label{fig:rxj0911b}}
\end{figure}

\begin{figure}
\epsscale{1.0}
\plotone{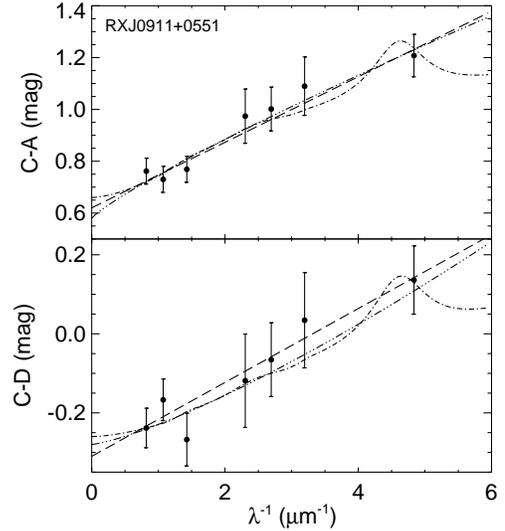}
\caption{RXJ0911+0551, C:  The upper panel shows the data points and the fits for the extinction of image C compared to image A.  The lower panel shows the corresponding plot for image C compared to image D.  The parameters of the fits can be seen in Table~\ref{tab:rxj0911c}.  See the caption of Figure \ref{fig:q2237c} for annotation overview.   \label{fig:rxj0911c}}
\end{figure}

We find relatively strong extinction in images B and C compared to images A and D.  Image D also shows some extinction when compared to A but the effect is consistent with zero within two sigmas.  We analyze the extinction curves of B and C compared to A and D.  The data points  and the fits can be seen in Figures~\ref{fig:rxj0911b} and \ref{fig:rxj0911c} and the parameters of the fits can be seen in Tables  \ref{tab:rxj0911b} and  \ref{tab:rxj0911c}. 

Assuming image A is completely unextinguished we can estimate the lower limit of the relative extinction of D compared to B and C, $E_D(B-V)/E_{B,C}(B-V)$.  For both B and C, we find that this ratio is around $1/3$ so we expect the extinction curve properties to be affected by both lines of sight (see discussion in \S~\ref{sec:extboth}).  That is, we do not expect the extinction curve we get from comparing B and C to D to represent the extinction curve along either line of sight unless their extinction properties are identical.  We see that the $R_V$ value of both B and C are lower when compared to image D than those we get from comparing them to image A suggesting that the extinction properties are indeed different (with image D having a higher $R_V$ value).  We note however that the values of $R_V$ do agree within one sigma for both the differential extinction curves for both B and C.

\subsubsection{HE0512$-$3329}
HE0512$-$3329 is a doubly imaged system first discovered by \citet{gregg} with an image separation of $0\farcs644$ and quasar redshift of $z=1.565$.  They estimated a redshift of $z=0.9319$ for the lensing object and found that the lens is most likely a spiral galaxy.  In addition, they estimate the differential reddening assuming negligible microlensing and a standard \cardelli~with $R_V=3.1$.  This yields $A(V)=0.34$ with the A image being redder than the B image.  \citet{wucknitz} worked further on disentangling microlensing and differential extinction, and estimated $A(V)= 0.07$ with A being the extinguished image.  This  fit results in an effective $R_V=-2.0$ which can be achieved if the two lines of sight have different $R_V$.  

\begin{deluxetable*}{lrrrrr} 
\tablecolumns{6} 
\tablewidth{0pc} 
\tablecaption{Extinction curve fit results for HE0512$-$3329} 
\tablehead{  \colhead{Extinction} & \colhead{$A(V)$}   & \colhead{$R_V$}    & \colhead{$\alpha$} & \colhead{$\Delta\hat{m}$}  & \colhead{$\chi^2_{\nu}$}}
\startdata 
\cardellitab & $ 0.14\pm0.04$ & $1.7\pm0.4$ &\nodata  & $-0.67\pm0.03$ &  $2.1\pm0.4$\\ 
Power law & $0.23\pm0.09$ & \nodata & $1.3\pm0.3$ & $-0.76\pm0.09$ & $1.4\pm0.4$ \\
Linear law & $0.35\pm0.02$ & \nodata & \textit{1.0} & $-0.86\pm0.04$ & $1.4\pm0.4$
\enddata 
\tablecomments{  The extinction of image A compared to image B.  Numbers quoted in italics were fixed in the fitting procedure.}
\label{tab:he0512}
\end{deluxetable*}

\begin{figure}
\epsscale{1.0}
\plotone{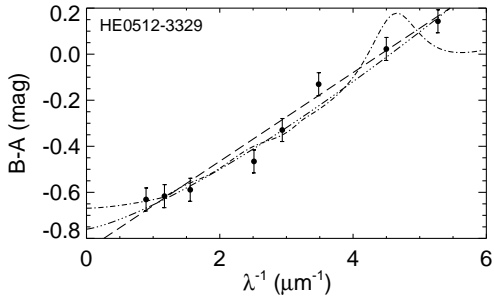}
\caption{HE0512$-$3329:   The plot shows the data points and the best fit to them.  The parameters of the fits can be seen in Table~\ref{tab:he0512}.  See the caption of Figure \ref{fig:q2237c} for annotation overview.  \label{fig:he0512}}
\end{figure}

In the case of HE0512$-$3329, it is the brighter image, A, which shows extinction with respect to the B image.  The system is interesting as one of the redshifted data points falls in the range where the $2175$ $\AA$ bump in the \cardelli~should lie (see Figure~\ref{fig:he0512}).  There is however no sign of a bump at $\lambda=2175$ $\AA$ and both the power law and the linear extinction law give a much better fit (see Table~\ref{tab:he0512} for the parameters of the fits).  We redo the fits with no constraints on the $R_V$ values to see if our data could be fit by a negative $R_V$ value but this does not change the result of $R_V=1.7\pm0.4$.  As there is no radio data available we do not constrain the intrinsic ratio in the fits and we can not constrain the microlensing signal.

Our results are not in agreement with those of \citet{wucknitz} who found that fits with the $2175$ $\AA$ bump better reproduced their data than those without, although the result was not highly significant.  In addition, they found that it is crucial to take microlensing into account when analyzing the extinction curve, which might explain the discrepancy.  However, the detected microlensing signal is only important at wavelengths lower than those we probe, with a small possible effect in the $B$- and $V$-bands.  Therefore, a microlensing signal consistent with the results of \citet{wucknitz} should not affect our results significantly.  In addition, we note that for their best fitting $R_V$ their fit curves downwards for $\lambda^{-1}<1$~$\mu m^{-1}$ which is not consistent with our measurement in the $K$-band (see Figure~\ref{fig:he0512}).

\subsubsection{MG0414+0534}
MG0414+0534 is a quadruply imaged system first discovered by \citet{hewitt} with image separation of up to $2^{\prime\prime}$.  The quasar, at redshift of $z=2.64$, shows evidence of being heavily reddened by dust in the lensing galaxy \citep{lawrence95}.  The lens, which has early type spectrum, is at redshift $z=0.9584$ \citep{tonry} and was modeled by \citet{falco1997} who found the brightness profile to be well represented by a de Vaucouleurs profile which is characteristic of an elliptical galaxy. \citet{falco1999} studied the extinction curve of this system and fitted it to a \cardelli~giving a best fit of $R_V=1.5$, assuming that all lines of sight have the same $R_V$.  \citet{angonin} studied the origin of the extinction and found, that while the differential extinction is likely due to the lensing galaxy, then there is also evidence for significant reddening which is intrinsic to the source.  \citet{katz} did an extensive radio survey of the system and found that there was no sign of variability in the radio flux ratios between their measurements and those of \citet{hewitt} except for the C/B image ratio.

\begin{figure}
\epsscale{1.0}
\plotone{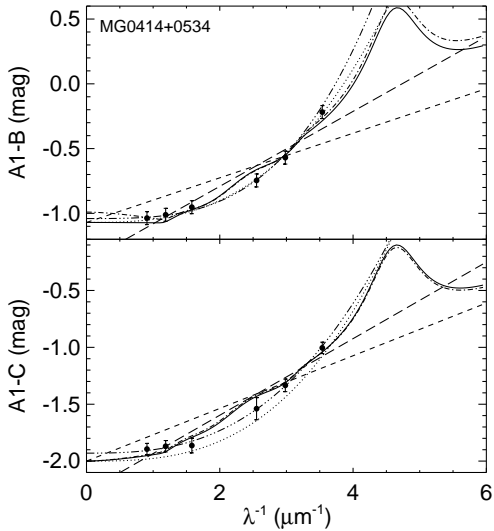}
\caption{MG0414+0534, A1:  The upper panel shows the data points and the fits for the extinction of image A1 compared to image B.  The lower panel shows the corresponding plot for image A1 compared to image C.  The parameters of the fits can be seen in Table~\ref{tab:mg0414a1}.  See the caption of Figure \ref{fig:q2237c} for annotation overview.  \label{fig:mg0414a1}}
\end{figure}

For MG0414+0534, components A1 and A2 show extinction when compared to images B or C with A2 being the more strongly extinguished image (see Table~\ref{tab:extinction}).  For the A1 image we find different effective extinction laws depending on whether we compare with image B or C (see Table~\ref{tab:mg0414a1} and Figure~\ref{fig:mg0414a1}).  In all cases the power law gives the best fit and the linear law the worst (we use the radio measurements of \citet{katz} to fix $\Delta\hat{m}$).  For the \cardelli~the $R_V$ values do not agree suggesting that perhaps the extinction of images B and C is affecting the differential extinction curve.  We also note though, that we would expect $A(V)$ for A1$-$B to be around 0.3, to be consistent with the other values in Table~\ref{tab:extinction}, but the best fitting values give a lower value.  We therefore perform another fit where we fix $A(V)=0.3$ in the fits for the A1$-$B pair and this gives $R_V=2.1\pm0.2;1.8\pm0.1$ for $\Delta\hat{m}$ fixed and free respectively, which are marginally consistent with the results compared to the C image.  However, the $\chi^2_\nu=2.6;2.0$ of these fits are significantly worse than those of the original fits.  We do not see any evidence for microlensing except in the case of SMC-like linear extinction which still results in a worse fit than the other two extinction laws.

\begin{figure}
\epsscale{1.0}
\plotone{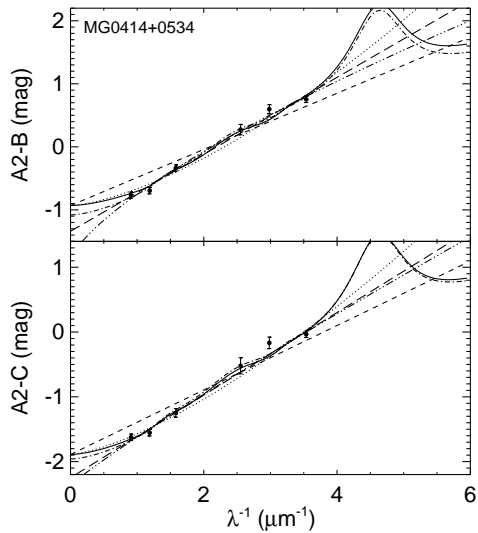}
\caption{MG0414+0534, A2:  The upper panel shows the data points and the fits for the extinction of image A2 compared to image B.  The lower panel shows the corresponding plot for image A2 compared to image C.  The parameters of the fits can be seen in Table~\ref{tab:mg0414a2}.  See the caption of Figure \ref{fig:q2237c} for annotation overview. \label{fig:mg0414a2}}
\end{figure}

For image A2 the \cardelli~gives the best fit when $\Delta\hat{m}$ is kept fixed but otherwise the different extinction laws give similar results (see Table~\ref{tab:mg0414a2} and Figure~\ref{fig:mg0414a2}).  The parameters of the \cardelli~are consistent when compared with images B and C suggesting that either A2 dominates the extinction signal or that B and C have similar extinction properties.  There is no evidence for microlensing except in the case of the linear extinction law.  We note that the absolute extinction of image A2, which must be greater or equal to the differential extinction in Table~\ref{tab:mg0414a2}, is very high given that the lens is an early type galaxy (\citet{goudfrooij1994} find $A(V)\lesssim0.35$ for their sample of early type galaxies).

\begin{deluxetable*}{lrrrrrrr} 
\tablecolumns{8} 
\tablewidth{0pc} 
\tablecaption{Extinction curve fit results for MG0414+0534: A1} 
\tablehead{ \colhead{Images} & \colhead{Extinction} & \colhead{$A(V)$}   & \colhead{$R_V$}    & \colhead{$\alpha$} & \colhead{$\Delta\hat{m}$}  &\colhead{$s$}& \colhead{$\chi^2_{\nu}$}}
\startdata 
A1$-$B & \cardellitab & $ 0.07\pm0.02$ & $0.4\pm0.1$ &\nodata  & $-0.99\pm0.03$ & \nodata& $1.5\pm0.5$\\ 
A1$-$B  & \cardellitab & $ 0.15\pm0.03$ & $0.9\pm0.2$ &\nodata  & \textit{-1.07} & \nodata&$1.6\pm0.5$\\

A1$-$C & \cardellitab & $ 0.29\pm0.04$ & $1.5\pm0.2$ &\nodata  & $-2.00\pm0.04$ &  \nodata&$1.7\pm0.5$\\ 
A1$-$C  & \cardellitab & $ 0.27\pm0.04$ & $1.4\pm0.2$ &\nodata  & \textit{-2.0} & \nodata&$1.5\pm0.4$\\

A1$-$B & Power law & $0.10\pm0.05$ & \nodata & $3.1\pm0.6$ & $-1.04\pm0.05$ & \nodata&$0.9\pm0.4$ \\
A1$-$B & Power law & $0.13\pm0.03$ & \nodata & $2.8\pm0.4$ & \textit{-1.07} & \nodata&$1.0\pm0.4$\\

A1$-$C& Power law & $0.15\pm0.07$ & \nodata & $2.7\pm0.6$ & $-1.93\pm0.05$ & \nodata&$1.0\pm0.4$ \\
A1$-$C& Power law & $0.24\pm0.04$ & \nodata & $2.1\pm0.3$ & \textit{-2.0} & \nodata&$1.1\pm0.4$\\

A1$-$B & Linear law & $0.53\pm0.04$ & \nodata & \textit{1.0} & $-1.37\pm0.05$ & \nodata&$2.2\pm0.5$\\
A1$-$B & Linear law & $0.31\pm0.02$ & \nodata & \textit{1.0} & \textit{-1.07} & \nodata&$3.4\pm0.4$\\
A1$-$B & Linear law & $0.49\pm0.03$ & \nodata & \textit{1.0} & \textit{-1.07} & $0.28\pm0.05$&$2.3\pm0.4$\\

A1$-$C & Linear law & $0.61\pm0.04$ & \nodata & \textit{1.0} & $-2.27\pm0.05$ & \nodata&$1.9\pm0.5$\\
A1$-$C & Linear law & $0.42\pm0.02$ & \nodata & \textit{1.0} & \textit{-2.0} & \nodata&$2.9\pm0.4$\\
A1$-$C & Linear law & $0.58\pm0.03$ & \nodata & \textit{1.0} & \textit{-2.0} & $0.26\pm0.05$&$1.9\pm0.5$

\enddata 
\tablecomments{  The extinction of image A1 compared to reference images B and C.  Numbers quoted in italics were fixed in the fitting procedure.}
\label{tab:mg0414a1}
\end{deluxetable*}

\begin{deluxetable*}{lrrrrrrr} 
\tablecolumns{8} 
\tablewidth{0pc} 
\tablecaption{Extinction curve fit results for MG0414+0534: A2 } 
\tablehead{ \colhead{Images} & \colhead{Extinction} & \colhead{$A(V)$}   & \colhead{$R_V$}    & \colhead{$\alpha$} & \colhead{$\Delta\hat{m}$}  & \colhead{$s$}& \colhead{$\chi^2_{\nu}$}}
\startdata 

A2$-$B  & \cardellitab & $ 0.87\pm0.05$ & $2.7\pm0.2$ &\nodata  & $-1.08\pm0.04$ & \nodata & $1.8\pm0.5$\\
A2$-$B  & \cardellitab & $ 0.69\pm0.03$ & $2.2\pm0.2$ &\nodata  & \textit{-0.93} & \nodata & $1.8\pm0.5$\\

A2$-$C & \cardellitab & $ 0.91\pm0.04$ & $2.6\pm0.1$ &\nodata  & $-1.96\pm0.04$ &  \nodata & $1.8\pm0.5$\\ 
A2$-$C  & \cardellitab & $ 0.81\pm0.04$ & $2.3\pm0.2$ &\nodata  & \textit{-1.89} & \nodata & $1.6\pm0.5$\\

A2$-$B & Power law & $1.6\pm1.2$ & \nodata & $0.7\pm0.2$ & $-1.8\pm1.2$ & \nodata &$1.7\pm0.5$ \\
A2$-$B & Power law & $0.65\pm0.03$ & \nodata & $1.49\pm0.07$ & \textit{-0.93} &\nodata & $2.3\pm0.5$\\

A2$-$C& Power law & $1.3\pm0.5$ & \nodata & $0.9\pm0.2$ & $-2.4\pm0.5$ & \nodata & $1.8\pm0.5$ \\
A2$-$C& Power law & $0.75\pm0.03$ & \nodata & $1.42\pm0.07$ & \textit{-1.89} & \nodata & $1.9\pm0.5$\\

A2$-$B & Linear law & $1.11\pm0.04$ & \nodata & \textit{1.0} & $-1.34\pm0.05$ & \nodata & $1.7\pm0.5$\\
A2$-$B & Linear law & $0.81\pm0.02$ & \nodata & \textit{1.0} & \textit{-0.93} & \nodata & $4.0\pm0.4$\\
A2$-$B & Linear law & $1.07\pm0.04$ & \nodata & \textit{1.0} & \textit{-0.93} & $0.39\pm0.05$ & $1.6\pm0.5$\\

A2$-$C & Linear law & $1.16\pm0.04$ & \nodata & \textit{1.0} & $-2.25\pm0.05$ & \nodata & $1.6\pm0.4$\\
A2$-$C & Linear law & $0.91\pm0.02$ & \nodata & \textit{1.0} & \textit{-1.89} & \nodata & $3.5\pm0.4$\\
A2$-$C & Linear law & $1.13\pm0.04$ & \nodata & \textit{1.0} & \textit{-1.89} & $0.35\pm0.05$ & $1.5\pm0.4$
\enddata 
\tablecomments{ The extinction of image A2 compared to reference images B and C.  Numbers quoted in italics were fixed in the fitting procedure.}
\label{tab:mg0414a2}
\end{deluxetable*}

\begin{deluxetable}{llll} 
\tablecolumns{4} 
\tablewidth{0pc} 
\tablecaption{MG0414+0534: The extinction properties of A2$-$A1} 
\tablehead{ \colhead{} & \colhead{$\Delta\hat{m}$}  &\colhead{$A(V)$} & \colhead{$R_V$}}
\startdata 
Fit & Free  & $0.61\pm0.11$ & $3.8\pm0.7$\\
Fit & Fixed & $0.53\pm0.03$ & $3.5\pm0.4$\\
C   & Free  & $0.62\pm0.06$ & $4.0\pm0.9$\\
C   & Fixed & $0.54\pm0.06$ & $3.4\pm1.0$\\
B   & Free  & $0.80\pm0.05$ & $5.4\pm2.5$\\
B   & Free  & $0.54\pm0.04$ & $3.7\pm1.3$
\enddata 
\tablecomments{  The first two lines give the results from the \cardelli~fit to the data.  The last four lines give the extinction properties calculated from eq. (\ref{eq:rdiff}) using the properties of A2 and A1 compared to images B and C from Tables~\ref{tab:mg0414a1} and \ref{tab:mg0414a2}.}
\label{tab:mg0414_rdiff}
\end{deluxetable}

As the extinction of A1 is significant compared to A2 we expect the extinction properties of both lines of sight to affect the A2$-$A1 extinction curve (see \S~\ref{sec:extboth}).  The fit of A2$-$A1 for the Galactic extinction curve gives us $R_V=3.5\pm0.4,3.8\pm0.7$ at $A(V)=0.53\pm0.03,0.61\pm0.11$ when $\Delta\hat{m}$ is kept fixed or free respectively.  If we assume that images B and C have zero extinction we can calculate the effective $R_V$ we expect to get from eq. (\ref{eq:rdiff}).  The results can be seen in Table~\ref{tab:mg0414_rdiff} and are in good agreement with the results of the fits.

The extinction of MG0414+0534 is high for an early type galaxy.  We can not exclude the possibility that the extinction may be due to an unknown foreground object and not the lensing galaxy itself.  Finally we note that our estimates of the differential extinction agree with those of \citet{falco1999} which were obtained by assuming standard Galactic extinction with $R_V=3.1$.

\subsubsection{MG2016+112}
\begin{deluxetable*}{lrrrrrrr} 
\tablecolumns{7} 
\tablewidth{0pc} 
\tablecaption{Extinction curve fit results for MG2016+112} 
\tablehead{ \colhead{Extinction} & \colhead{$A(V)$}   & \colhead{$R_V$}    & \colhead{$\alpha$} & \colhead{$\Delta\hat{m}$} & \colhead{$s$} & \colhead{$\chi^2_{\nu}$}}
\startdata 
\cardellitab & $ 0.20\pm0.03$ & $3.0\pm0.5$ &\nodata  & \textit{-0.092}& \nodata & $1.7\pm0.4$\\  
\cardellitab & $0.1\pm0.1$ & $1.8\pm1.0$ & \nodata & $-0.01\pm0.10$ & \nodata & $1.9\pm0.5$\\ 
\cardellitab & $0.11\pm0.09$ & $1.8\pm 1.3$ & \nodata & \textit{-0.092} & $-0.1\pm0.1$ & $1.8\pm0.5$\\ 
Power law  & $0.18\pm0.03 $ & \nodata & $1.4\pm0.2$ & \textit{-0.092} & \nodata & $1.4\pm0.4$\\
Power law  & $0.11\pm0.09$   & \nodata & $1.8\pm0.6$ & $-0.02\pm0.10$ & \nodata & $1.5\pm0.5$\\  
Power law  & $0.12\pm0.09$ & \nodata & $1.7\pm0.5$ & \textit{-0.092} & $-0.06\pm0.11$ & $1.5\pm0.5$\\  
Linear law& $0.23\pm0.01$ & \nodata & \textit{1.0} & \textit{-0.092} & \nodata & $1.6\pm0.4$\\  
Linear law& $0.28\pm0.03$ & \nodata & \textit{1.0} & $-0.18\pm0.06$ &\nodata & $1.6\pm0.5$\\  
Linear law& $0.27\pm0.03$ & \nodata & \textit{1.0} & \textit{-0.092} & $0.08\pm0.06$ & $1.6\pm0.5$\\
\enddata 
\tablecomments{  The extinction of image B compared to reference image A.  Numbers quoted in italics were fixed in the fitting procedure.}
\label{tab:mg2016}
\end{deluxetable*}

MG2016+112 was discovered by \citet{lawrence} and has a giant elliptical lensing galaxy at redshift $z=1.01$ \citep{schneider85, schneider86}.  The system consists of two images, A and B, of the quasar at redshift $z=3.273$ and an additional image C which may be a third image of the quasar with an additional signal from another galaxy and has been challenging to model \citep{lawrence, lawrence93,nair}.  The flux of images A and B in the radio at 5 GHz was determined by \citet{garrett} to be $15.8$~mJy and $17.2$~mJy respectively.

\begin{figure}
\epsscale{1.0}
\plotone{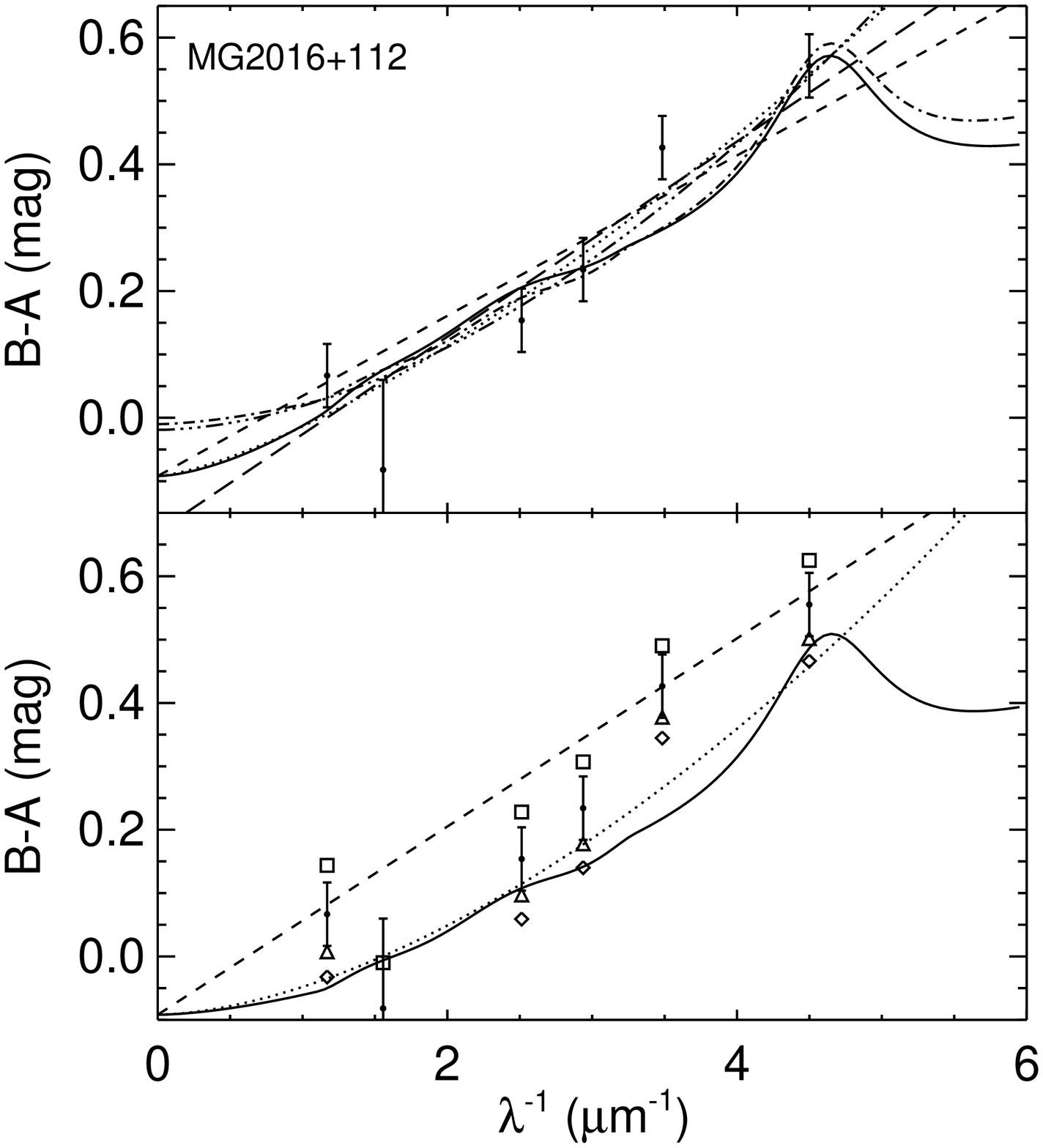}
\caption{MG2016+112:   The upper panel shows the data points and the best fit extinction curves as given in Table~\ref{tab:mg2016}.  The lower panel shows the original data points and their shift due to a microlensing signal. The parameters of the fits can be seen in Table~\ref{tab:mg2016}.  See the caption of Figure \ref{fig:q2237c} for annotation overview.\label{fig:mg2016}}
\end{figure}

This is the highest redshift system in our sample, and is also interesting since one of the data points lands in the range where the $2175$ $\AA$ bump in the \cardelli~should be (see Figure~\ref{fig:mg2016}).  However, the extinction signal is very weak with $A(V)=0.1\pm0.1,0.11\pm0.09,0.28\pm0.03$ at $\chi^2_\nu=1.9,1.5,1.6$ for the \cardelli, power law and linear law respectively (see Table~\ref{tab:mg2016} for the parameters of the fits).  When we fix $\Delta\hat{m}$ we find somewhat higher extinction of $A(V)=0.20\pm0.03,0.18\pm0.03,0.23\pm0.01$ at $\chi^2_\nu=1.7,1.4,1.6$.  In both cases a power law or a linear law is marginally preferred to a \cardelli.  We also analyze the data with respect to a possible microlensing signal but only find a weak microlensing signal (see Table~\ref{tab:mg2016} and Figure~\ref{fig:mg2016}).  Finally we note that our results for the \cardelli~are consistent with the results of \citet{falco1999}.

\subsection{The full sample}
\label{sec:res_full}

In this section we study the properties of the sample as a whole.  We look for correlations between various parameters and, in particular, search for any dependence on the redshift or the morphology of the galaxies.  Furthermore, we discuss the low $R_V$ values found in SN Ia studies and the possible complementarity of lensing extinction curve studies.  We study, on the one hand, a `golden sample' and, on the other hand, we analyze the full sample.  The `golden sample' is defined to include the image pair with the strongest differential extinction for each lens.  In addition Q0142$-$100 is excluded from the `golden sample' (see \S \ref{sec:q0142}).  The `golden sample' therefore consists of eight pairs of images, of which seven have strong enough extinction to analyse the extinction curve.  If not otherwise stated, the results apply to the full sample.

\subsubsection{$A(V)$ as a function of distance from center of the lensing galaxy}
\begin{figure}
\epsscale{1.0}
\plotone{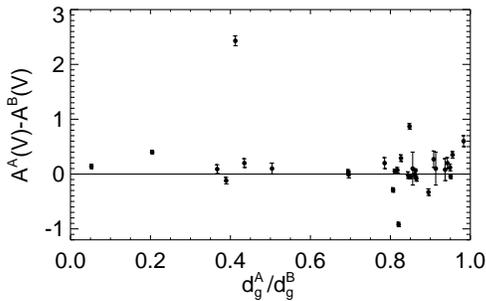}
\caption{The differential $A(V)$ for a pair of images~vs.~the ratio of the distances from the center of the lensing galaxy.  The differential $A(V)$ is defined as negative if the image closer to the galaxy is less extinguished.  The figure shows that images closer to the galaxy tend to be the more extinguished but that when the ratio approaches $1$ the scatter increases. \label{fig:A_dist}}
\end{figure}

To study the distribution of $A(V)$ as a function of distance from the lens galaxy, we analyze the sample using two methods.  First we plot, in Figure~\ref{fig:A_dist}, the differential $A(V)$ of the image pairs  (from Table~\ref{tab:extinction}) as a function of the ratio of the distances from the center of the galaxy (from Table~\ref{tab:lenses}).  We assign a negative value to the $A(V)$, in those cases where the more distant image is the more strongly extinguished one.  One can see, that when the ratio is small, the image which is nearer the center of the galaxy is the more extinguished one.  However, when the ratio approaches one, the $A(V)$ becomes more evenly scattered around zero.

\begin{figure}
\epsscale{1.0}
\plotone{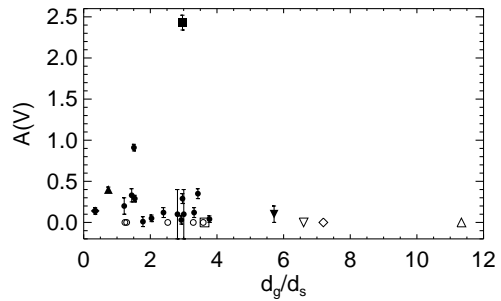}
\caption{$A(V)$ as a function of the distance of the image relative to the scale radius of the lensing galaxy.  We assign an absolute $A(V)$ to the images by assuming that the least extinguished image for each system is non-extinguished.  The quads are symbolized as circles and the four doubles are symbolized as triangles (up and down facing), a diamond and a box.  The non-extinguished reference images are marked on the plot by open symbols. The error bars for $d_g/d_s$ are smaller than the plotted symbols.  We see that the $A(V)$ values mostly lie in the range of $0-0.5$ for $d_g/d_s \lesssim 4$ and drop for higher values.\label{fig:A_scale}}
\end{figure}

For the second method, we assume that the image with the weakest extinction signal is indeed non-extinguished.  We define an absolute $A(V)$ for the other images, by taking the differential extinction compared to this reference image, which we plot as a function of distance from the center of the galaxy, scaled by the lens galaxy scale length\footnote{The scale length is taken to be the effective radius of a de Vaucouleurs profile fit from \citet{rusin2003}.} (see  Figure~\ref{fig:A_scale} and Table~\ref{tab:lenses}).  From the plot we can see that most $A(V)$ values lie in the range of $0$--$0.5$ for distances smaller than around four scale lengths, but drop for more distant images.

Both of these results are consistent with the expectation that the more distant image is on average more likely to pass outside the galaxy and thus not be affected by extinction.  When the distances become similar, secondary effects due to the non-symmetric shape of the lens start becoming important, creating a scatter in the  $A(V)$~vs.~distance plots.  This is in particular the case for the quads where the distances tend to be similar.

\subsubsection{The different extinction laws}
We investigate whether our sample shows a preference for one type of extinction law to another and whether the type of extinction depends on the galaxy type.  We also study the correlation between the parameters of the different fits.

We find that when $\Delta\hat{m}$ is allowed to vary, our sample does not show a preference for one extinction law over the other (the mean of the $\chi^2_\nu$ is $\bar{\chi}^2_\nu=1.7,1.6,1.8$ for the \cardelli, power law and linear law) although individual systems can show a strong preference.  If we alternatively look at the fits where $\Delta\hat{m}$ was fixed we see that the power law and \cardelli~are preferred over the linear law in the sample as a whole (with  $\bar{\chi}^2_\nu=2.7,2.1,4.3$) but again individual systems can show different behaviors.

There are three late type galaxies in our sample.  One (HE0512$-$3329) shows a clear preference for an SMC linear law extinction, one (B1152+199) shows a preference for a \cardelli~and the third (Q2237+030) gives equally good fits to all the extinction laws  (which is expected due to its low redshift, see \S~\ref{sec:pureext}).  For the early type galaxies there is also no clear preference for one type of extinction law.  Three systems (PG1115+080, Q0142$-$100, RXJ0911+0551) show no preference for one extinction law over the other, one (MG0414+0534) favors a power law with power index $\alpha=2-3$ for one of the images (which may be affected by extinction along both lines of sight) and one (MG0216+112) shows a weak preference for a power or a linear law over the \cardelli.  We therefore conclude that there is no evidence for a correlation between galaxy type and type of extinction in our sample.

When confining the analysis to a Galactic extinction, we find that the mean $R_V$ (for the `golden sample') of the late type galaxies ($\bar{R}_V^{late}=2.3\pm0.5$) is marginally lower than that of the early type galaxies ($\bar{R}_V^{early}=3.2\pm0.6$), however they are consistent within the error bars and the difference may be due to low number statistics.  This is further discussed in \S \ref{sec:lowRv}.

\begin{figure}
\epsscale{1.0}
\plotone{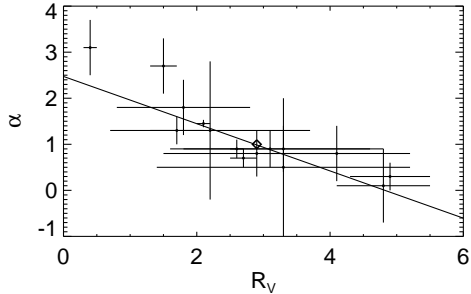}
\caption{The power index $\alpha$~vs.~$R_V$.  The data points consist of all image pairs where the extinction curves were analyzed (from fits with $\Delta\hat{m}$ free). The plot shows a clear correlation between $\alpha$ and $R_V$, with lower $\alpha$ giving higher $R_V$, which is consistent with lower $R_V$ giving a steeper rise into the UV.  The point corresponding to $\alpha=1.0$ (SMC type extinction) and $R_V=2.9$ \citep[the mean $R_V$ for the SMC, as determined by][]{pei2}, is marked by a diamond.\label{fig:alpha_R}}
\end{figure}

\begin{figure}
\epsscale{1.0}
\plotone{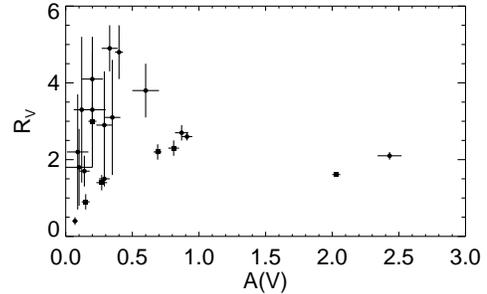}
\caption{The figure shows $R_V$~vs.~$A(V)$.  The data points consist of all image pairs where the extinction curves were analyzed (filled circles from fits with $\Delta\hat{m}$ free and filled boxes from fits with $\Delta\hat{m}$ fixed). The figure shows, as expected, no correlation between $R_V$ and the amount of extinction.\label{fig:R_A}}
\end{figure}

We also study the correlation of the parameters $R_V$ and $\alpha$ for each system to demonstrate the consistency of the two approaches.  As expected, we find that there is a strong correlation with larger $R_V$ giving smaller $\alpha$, as seen in Figure~\ref{fig:alpha_R}, consistent with smaller $R_V$ giving steeper rise into the UV.  The exact relationship between $R_V$ and $\alpha$ can be derived by solving equations (\ref{eq:car}) and (\ref{eq:alpha}) and is wavelength dependent.  A first order linear fit to our data gives $\alpha=(2.5\pm0.2)-(0.51\pm0.09)R_V$ which is an applicable approximation within the wavelength range of our data.  In addition we check whether the strength of the extinction is correlated to $R_V$, but we find no evidence for such a correlation (see Figure~\ref{fig:R_A}).

\begin{figure}
\epsscale{1.0}
\plotone{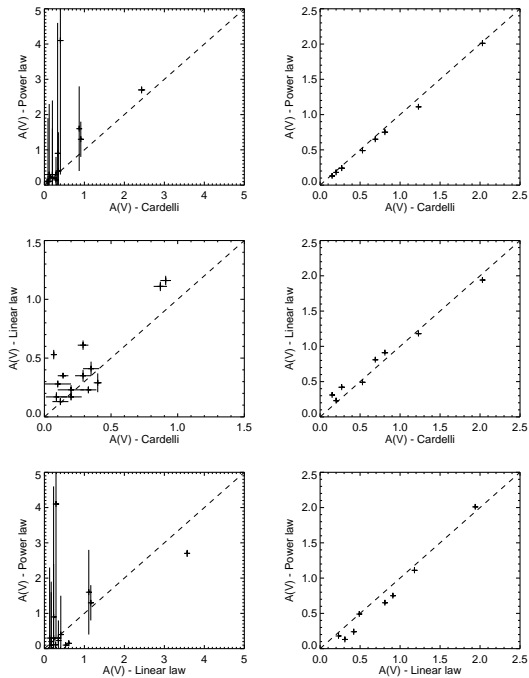}
\caption{The graphs show the correlation between the values of $A(V)$ derived for the different fits.  The left column shows the distribution for the fits where $\Delta\hat{m}$ is a fitted parameter and the right column gives the corresponding distribution when $\Delta\hat{m}$ is fixed.  The dashed line corresponds to $x=y$, which would correspond to perfect agreement in $A(V)$ between the fits, and is plotted for reference. \label{fig:A_A}}
\end{figure}

Finally we study the correlation between the values of $A(V)$ which were found corresponding to the different extinction laws.  The results can be seen in Figure~\ref{fig:A_A}.  We see that when $\Delta\hat{m}$ is free, the power law favors higher $A(V)$ than the other two extinction laws (see left bottom and top panels in Fig.~\ref{fig:A_A}).  When $\Delta\hat{m}$ is fixed, the correspondence becomes much better with the power law giving marginally lower values.  The agreement between the $A(V)$ values derived for the \cardelli~and the linear law are in general good, regardless of whether we keep $\Delta\hat{m}$ fixed or free (with the exception that for $A(V)\gtrsim1$ the linear law gives higher results when $\Delta\hat{m}$ is free).

Figure~\ref{fig:A_A} also clearly demonstrates that the $A(V)$ values become much better constrained when $\Delta\hat{m}$ is fixed in the fitting.  This suggests that it would be valuable for a future extinction survey, to do a simultaneous radio survey for the systems, in order to constrain the intrinsic magnitude ratio of the images.

\subsubsection{Evolution with redshift}
\label{sec:redshift}
\begin{figure}
\epsscale{0.9}
\plotone{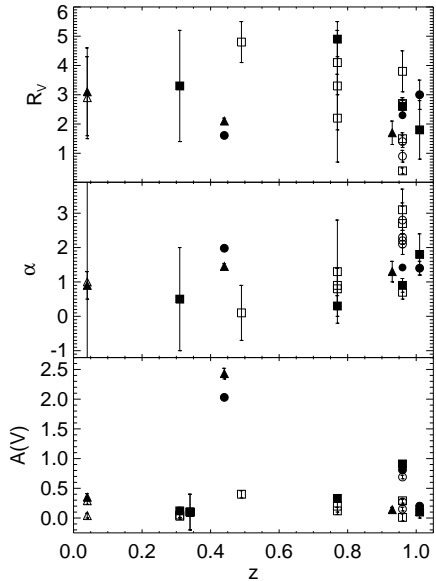}
\caption{ The top panel shows $R_V$ as a function of $z$.  The middle panels shows $\alpha$~vs.~$z$.  The bottom panels shows $A(V)$ (as given by a \cardelli~fit)~vs.~$z$.  On all panels, triangles denote late type galaxies and boxes denote early type galaxies where the values are taken from the fits with $\Delta\hat{m}$ kept free.  Circles denote the corresponding fits where  $\Delta\hat{m}$ was fixed.  Filled symbols correspond to a `golden' sample as defined in \S \ref{sec:res_full}.  We see no strong evolution with $z$ but lower values of $R_V$ seem to appear at higher $z$.  \label{fig:all_z}}
\end{figure}

Next, we investigate the behavior of our sample as a function of redshift.  The plots of $R_V$ and $\alpha$ as a function of redshift can be seen in the upper two panels of Figure~\ref{fig:all_z}.  We do not see any strong correlation, in either the full nor the `golden sample', although the lower values of $R_V$ and the higher values of $\alpha$ seem to appear at higher $z$.  Our results do not confirm evolution of $R_V$ with redshift, with lower $R_V$ at higher $z$, as suggested by \citet{ostman}, but do not exclude such evolution either.  We stress that a larger sample would be needed to make any conclusive claims.

We also investigate whether there might be an evolution in the amount of dust extinction with redshift.  Again, we use two samples, the `golden sample' as defined above, and a sample consisting of the highest differential extinction deduced for each image.  The resulting plot can be seen in the bottom panel of Figure~\ref{fig:all_z} and does not show any correlation between $A(V)$ and $z$.

\subsubsection{Low $R_V$ values and Type Ia SNe}
\label{sec:lowRv}
Recent studies of SNe Ia have suggested that $R_V$  values for SN hosts \citep[which are mostly late type, see][]{sullivan} could be lower than those for the Milky Way \citep[see e.g.,][]{riess1996b, krisciunas, wang2006} suggesting that SN hosts are systematically different from the Milky Way.  \citet{wang} however suggest that the reason for the low values of $R_V$ might be due to circumstellar dust around the SNe themselves.  The presence of such dust would cause inaccurate estimates of the dust extinction of the host galaxies of the SNe Ia.  

Lensing studies have also seen more extreme $R_V$ values than those in the Milky Way \citep[see e.g.,][]{motta, wucknitz} but this has been criticized as possibly being due to extinction along both lines of sight.  However, by choosing systems where the extinction of the measured image dominates the extinction of the reference image, this effect can be avoided (see discussion in \S~\ref{sec:extboth}).  In our sample the mean $R_V$ value is $\bar{R}_V=2.8\pm0.3$ (with RMS scatter of $1.2$) for the full sample, and $\bar{R}_V=2.8\pm0.3$ (with RMS scatter of $1.1$) for the `golden sample'.  These values are marginally lower than, but consistent with, the Milky Way mean value of $R_V=3.1$.  If we look at $R_V$ for the late and early type galaxies separately, we find $\bar{R}_V^{early}=3.2\pm0.6$ and $\bar{R}_V^{late}=2.3\pm0.5$ for the `golden' sample.  These values are consistent with each other within the quoted error bars, but it is interesting that the late type galaxies have a lower mean $R_V$, in agreement with SN Ia studies.  A larger sample would be needed to determine whether this is a real trend, or due to low number statistics.

The mean extinction in our  `golden sample' is $\bar{A}(V)=0.56\pm0.04$ (with RMS scatter of $0.80$).  This gives a lower limit to the mean absolute extinction, as the mean $A(V)$ value is lowered if the reference image is also extinguished.  If we remove the highly extinguished system B1152+199 from our sample, the mean of the `golden sample' reduces to $\bar{A}(V) = 0.29\pm0.05$ (with RMS scatter of $0.29$).  If we instead take the highest differential extinction for each image for the full sample into account we get $\bar{A}(V) = 0.33\pm0.03$ (with RMS scatter of $0.56$ (or $\bar{A}(V) = 0.21\pm0.03$ without B1152+199).  All these values are high enough to cause systematic effects in the calibration of SNe Ia.

A lower mean $R_V$ from a lensing study would strengthen the results of low $R_V$ values from SN studies applying to the interstellar medium.  A lower real $R_V$ value than the assumed one would lead to an overestimation of $A(V)$ given a measurement of $E(B-V)$ (as $A(V)=R_V E(B-V)$).  It is therefore important that the extinction properties of higher redshift galaxies, and SN hosts in particular, be further investigated as assuming a mean Galactic extinction with $R_V=3.1$ in the analysis of SNe Ia could affect the cosmological results.

Lensing galaxies and SN Ia hosts are distributed over a similar redshift range (from $z=0$ to $z\sim1$) and consist of both early type and late type galaxies.  The majority of lensing galaxies are massive early type galaxies \citep{kochanek} whereas the SN Ia hosts are mostly late type \citep{sullivan}.  It would however be possible to select sub-samples of either group which would have the same morphology distribution as the other.  Therefore, studies of the extinction properties of lensing galaxies can complement future dark energy SN Ia surveys, providing an independent measurement of the extinction properties of the SN Ia type hosts.

\section{Summary}
\label{sec:summary}
We have presented an imaging survey of the extinction properties of 10 lensing galaxies using multiply imaged quasars observed with the ESO VLT in the optical and the NIR.  We have made a dedicated effort to reduce the number of unknowns and effects which can mimic extinction.  We have explored, analytically and in simulations, the effects of extinction along both sight lines.  We find that it is not crucial for the reference image to have zero extinction, as long as its extinction is small compared to the other image.  We also study the effects of achromatic microlensing and find that to account for such an effect in photometric data, it is crucial to have constraints on the intrinsic magnitude difference of the images.

We were able to study the extinction of 9 out of 10 of the systems in the survey, the last one had to be discarded due to contamination by the lensing galaxy.  Out of the 9 systems, 8 have a two sigma extinction signal for at least one image pair, which was our limit for doing further extinction curve analysis. However, we suspect that one of those is also contaminated by the lensing galaxy (Q0142$-$100) and exclude it from our  `golden sample'.  The mean extinction for the `golden sample' is $\bar{A}(V)=0.56\pm0.09$, using \cardelli~parametrization, and the mean $R_V$ is $\bar{R}_V=2.8\pm0.4$ (compared to $\bar{R}_V=2.8\pm0.3$ for the full sample), which is consistent with the mean $R_V=3.1$ found for the Galaxy.  The systems show various extinction properties.  There is no strong evidence for a correlation between morphology and extinction properties.  As our sample covers a broad range in redshifts ($z=0.04$--$1.01$) we have also looked for evolution with redshift.  However, our results neither confirm nor refute evolution of extinction parameters with redshift and we stress that a larger sample would be needed to make any conclusive claims.

Finally we wish to point out that large studies of gravitationally lensed quasars are ideal to study the possible evolution of extinction properties as they are spread over a redshift range from $z=0$ to $z\approx1$.  Furthermore, the quasars do not affect the environment of the galaxy we wish to study as is the case in SN Ia studies.  For further improvements, however, higher resolution and deeper imaging for a larger sample would be required making a dedicated study with space based telescopes of considerable interest.  A simultaneous radio survey, in order to constrain the intrinsic ratio of the images, would also further improve the results.  Such a study could complement future dark energy SN Ia surveys, providing an independent measurement for the extinction properties of SN Ia type host galaxies.

\acknowledgments

The Dark Cosmology Centre is funded by the Danish National Research Foundation.  This work was supported by the European Community's Sixth Framework Marie Curie Research Training Network Programme, Contract No. MRTN-CT-2004-505183 "ANGLES".  ST acknowledges support from the Danish Natural Science Research Council.

\clearpage

\begin{deluxetable}{lllll}
\tablecolumns{5}
\tablewidth{0pc}
\tablecaption{Overview of extinction properties for the \cardelli~fit.}
\tablehead{ \colhead{Lens} & \colhead{Lens redshift} & \colhead{Image pair}   & \colhead{$A(V)$}    & \colhead{$R_V$}}
\startdata
Q2237+030 & 0.04 & B$-$A	 & $0.04\pm0.04$ & \\
          & & C$-$A & $0.29\pm 0.06$ & $2.9 \pm 1.4$  \\
          & & D$-$A &  $0.35\pm0.06$ & $3.1\pm1.5$ \\ 
          & & C$-$B &  $0.2\pm0.1$& \\ 
          & & D$-$B &   $0.27\pm0.15$&\\ 
          & & D$-$C &   $0.08\pm0.05$& \\
PG1115+080 & 0.31 &A2$-$A1 & $0.12\pm 0.06$ & $3.3 \pm 1.9$ \\ 
          & & B$-$A1 & $0.05\pm0.04$ &\\ 
          & & C$-$A1 & $0.03\pm0.05$ &  \\ 
          & & A2$-$B & $0.05\pm0.03$ & \\ 
          & & A2$-$C & $0.07\pm0.05$ & \\ 
          & & B$-$C & $0.03\pm0.05$ & \\
B1422+231 & 0.34 & B$-$A 	 & $0.08\pm0.20$ &  \\
          & & A$-$C & $0.1\pm0.3$ &\\ 
          & & D$-$A & \nodata & \\ 
          & & B$-$C & $0.1\pm0.3$ &\\
          & & D$-$B & \nodata & \\ 
          & & D$-$C & \nodata& \\ 
B1152+199 & 0.44 & B$-$A   & $2.43\pm0.09$ & $2.1 \pm 0.1$  \\
Q0142$-$100 & 0.49 & B$-$A	 & $0.40\pm0.03$ & $4.7\pm0.7$\\ 
B1030+071 & 0.60  & B$-$A	 & \nodata &  \\ 
RXJ0911+0551 & 0.77 & B$-$A 	 & $0.33\pm0.06$& $4.9\pm0.6$\\  
          & & C$-$A & $0.20\pm0.10$ & $3.3\pm1.5$\\ 
          & & D$-$A &$0.12\pm0.06$ &  \\ 
          & & B$-$C & $0.02\pm0.06$ &  \\ 
          & & B$-$D & $0.20\pm0.08$ & $4.1\pm1.1$ \\ 
          & & C$-$D & $0.09\pm0.08$ & $2.2\pm1.5$\\ 
HE0512$-$3329 & 0.93 & A$-$B	 & $0.14 \pm 0.04$&$1.7\pm 0.4$\\
MG0414+0534 & 0.96 & A2$-$A1 	 & $0.6\pm0.1$&  $3.8 \pm 0.7$  \\  
          & & A1$-$B & $0.07\pm0.02$  & $0.4\pm 0.1$ \\ 
          & & A1$-$C & $0.29\pm0.04$ &$1.5\pm0.2$ \\
          & & A2$-$B & $0.87\pm0.05 $ & $2.7\pm0.2$\\ 
          & & A2$-$C &$0.91\pm0.04$ & $2.6\pm0.1$ \\
          & & B$-$C & $0.01\pm0.06$ &  \\ 
MG2016+112 & 1.01 & B$-$A	 &$0.1\pm0.1$ & $1.8\pm1.0$ \\ 
\enddata
\label{tab:extinction}
\tablecomments{Table of extinction properties for the 10 lensing systems.  The systems are ordered according to increasing redshift.  The fit results are from the \cardelli~with $\Delta\hat{m}$ free.}
\end{deluxetable}


\begin{thebibliography}{}

\bibitem[Alcalde(2002)]{alcalde} Alcalde, D. et al. 2002, ApJ, 572, 729

\bibitem[Angonin-Willaime et al.(1999)]{angonin} Angonin-Willaime, M.-C., Vanderriest, C., Courbin, F., Burud, I., Magain, P., \& Rigaut, F. 1999, AA, 347, 434

\bibitem[Bade et al.(1997)]{bade} Bade, N., Siebert, J., Lopez, S., Voges, W., \& Reimers, D. 1997, A\&A, 317, L13

\bibitem[Bechtold \& Yee(1995)]{bechtold} Bechtold, J. \& Yee, H. K. C. 1995, AJ, 110, 1984

\bibitem[Bertin \& Arnouts(1996)]{bertin} Bertin, E. \& Arnouts, S. 1996, AA, 117, 393

\bibitem[Bianchi et al.(1996)] {bianchi} Bianchi, L., Clayton, G. C., Bohlin, R. C., Hutchings, J. B., \& Massey, P. 1996, ApJ, 471, 203

\bibitem[Burud et al.(1998a)]{burud} Burud, I. et al. 1998a, ApJ, 501, L5

\bibitem[Burud et al.(1998b)]{burudb} Burud, I., Stabell, R., Magain, P., Courbin, F., \O stensen, R., Refsdal, S., Remy, M., \& Teuber, J. 1998b, A\&A, 339, 701

\bibitem[Cardelli et al.(1989)] {cardelli} Cardelli, J. A., Clayton, G.C. \& Mathis, J.S. 1989, ApJ, 345, 245

\bibitem[Christian et al.(1987)]{christian} Christian, C. A., Crabtree, D., \& Waddell, P. 1987, ApJ, 312, 45

\bibitem[Corrigan et al.(1991)]{corrigan} Corrigan, R. T. et al. 1991, AJ, 102, 34

\bibitem[Devillard(1997)]{devillard} Devillard, N, 1997, `The eclipse software', The Messenger, 87, 19

\bibitem[Ellison et al.(2005)]{ellison} Ellison, S. L., Hall, P. B., \& Lira, P. 2005, AJ, 130, 1345

\bibitem[Falco et al.(1996)]{falco96} Falco, E. E., Leh\'ar, J., Perley, R. A., Wambsganss, J, \& Gorenstein, M. V. 1996, AJ, 112, 897

\bibitem[Falco et al.(1997)] {falco1997} Falco, E. E., Leh\'ar, J., \& Shapiro, I. I. 1997, AJ, 113, 540

\bibitem[Falco et al.(1999)]{falco1999} Falco, E. E. et al. 1999, ApJ, 523, 617

\bibitem[Fassnacht \& Cohen(1998)]{fassnacht} Fassnacht, C. D. \& Cohen, J. G. 1998, AJ, 115, 377

\bibitem[Fitzpatrick et al.(1999)]{fitzpatrick} Fitzpatrick, E. L. 1999, PASP, 111, 63

\bibitem[Garrett et al.(1994)]{garrett} Garrett, M. A., Muxlow, T. W., Patnaik, A. R., \& Walsh, D. 1994, MNRAS, 269, 902

\bibitem[Gil-Merino et al.(2005)]{gil-merino} Gil-Merino, R., Wambsganss, J., Goicoechea, L. J., \& Lewis, G. F. 2005, A\&A, 432, 83

\bibitem[Goicoechea et al.(2005)]{goicoechea} Goicoechea, L. J, Gil-Merino, R. \& Ull\'an, A. 2005, MNRAS, 360, L60

\bibitem[Goudfrooij(1994)]{goudfrooij1994} Goudfrooij, P., de Jong, T., Hansen, L., \& N\o rgaard-Nielsen, H. U. 1994, MNRAS, 271, 833

\bibitem[Goudfrooij(2000)]{goudfrooij} Goudfrooij, P. 2000, ASP Conf. Ser. 209, Small Galaxy Groups, ed. M. Valtonen, \& C. Flynn (San Francisco:ASP), 74

\bibitem[Gregg et al.(2000)]{gregg} Gregg, M. D., Wisotzki, L., Becker, R. H., Maza, J., Schechter, P. L., White, R. L., Brotherton, M. S., \& Winn, J. N.  2000, AJ, 119, 2535

\bibitem[Hege et al.(1981)]{hege} Hege, E. K., Hubbard, E. N., Strittmatter, P. A., \& Worden, S. P. 1981, ApJ, 248, L1

\bibitem[Hewitt et al.(1992)] {hewitt} Hewitt, J. N., Turner, E. L., Lawrence, C. R., Schneider, D. P., \& Broody, J. P. 1992, AJ, 104, 968

\bibitem[Hjorth et al.(2002)]{hjorth} Hjorth, J. et al. 2002, ApJ, 572, L11

\bibitem[Huchra et al.(1985)]{huchra} Huchra, J., Gorenstein, M., Kent, S., Shapiro, I., \& Smith, G. 1985, AJ, 90, 691

\bibitem[Irwin et al.(1989)]{irwin} Irwin, M. J., Webster, R. L., Hewett, P. C., Corrigan, R. T., \& Jedrzejewski, R. I. 1989, AJ, 98, 1989

\bibitem[Jakobsson et al.(2004)]{palli} Jakobsson, P. et al. 2004, A\&A, 427, 785

\bibitem[Jaunsen et al.(1997)] {jaunsen} Jaunsen, A. O. \& Hjorth, J. 1997, A\&A, 317, L39

\bibitem[Kann et al.(2006)]{kann} Kann, D. A., Klose, S., \& Zeh. A. 2006, ApJ, 641, 993

\bibitem[Katz \& Hewitt(1993)]{katz} Katz, C. A., \& Hewitt, J.N. 1993, ApJ, 409, L9

\bibitem[Kneib et al.(2000)]{kneib} Kneib, J.-P., Cohen, J. G., \& Hjorth, J. 2000, ApJ, 544, L35

\bibitem[Kochanek et al.(2000)]{kochanek} Kochanek, C. S. 2000, ApJ, 543, 131

\bibitem[Kochanek(2004)]{kochanek2004} Kochanek, C. S. 2004, ApJ, 605, 58

\bibitem[Krisciunas et al.(2000)]{krisciunas} Krisciunas, K., Hastings, N. C., Loomis, K., McMillan, R., Rest, A., Riess, A. G., \& Stubbs, C. 2000, AJ, 539, 658

\bibitem[Kristian et al.(1993)] {kristian} Kristian, J. et al. 1993, AJ, 106, 1330

\bibitem[Krolik(1999)] {krolik} Krolik, J. H. 1999, Active Galactic Nuclei, Princeton University Press, New Jersey

\bibitem[Kundi\'c et al.(1997a)]{kundica} Kundi\'c, T., Cohen, J. G., Blandford, R. D., \& Lubin, L. M. 1997, AJ, 114, 507

\bibitem[Kundi\'c et al.(1997b)]{kundic} Kundi\'c, T., Hogg, D. W., Blandford, R. D., Cohen, J. G., Lubin, L. M., \& Larkin, J. E. 1997, AJ, 114, 2276

\bibitem[Lawrence et al.(1984)]{lawrence}  Lawrence, C. R., Schneider, D. P., Schmidt, M., Bennet, C. L., Hewitt, J. N., Burke, B. F., Turner, E. L, \& Gunn, J. E. 1984, Science, 223, 46

\bibitem[Lawrence et al.(1992)]{lawrence92} Lawrence, C. R., Neugebauer, G., Weir, N., Matthews, K., \& Patnaik, A. R. 1992, MNRAS, 259, 5p

\bibitem[Lawrence et al.(1993)]{lawrence93}Lawrence, C. R., Neugebauer, G, \& Matthews, K. 1993, AJ, 105, 17

\bibitem[Lawrence et al(1995)]{lawrence95} Lawrence, C. R., Elston, R., Januzzi, B. T., \& Turner, E. L. 1995, AJ, 110, 2570 

\bibitem[Leh\'ar et al.(2000)]{lehar} Leh\'ar et al. 2000, ApJ, 536, 584

\bibitem[MacAlpine \& Feldman(1982)]{macalpine} MacAlpine, G. M. \& Feldman, F. R. 1982, ApJ, 261, 412

\bibitem[Madau et al.(1998)] {madau} Madau, P., Pozzetti, L. \& Dickinson, M. 1998, ApJ, 498, 106

\bibitem[Magain et al.(1998)]{magain} Magain, P., Courbin, F., \& Sohy, S. 1998, ApJ, 494, 472

\bibitem[Malhotra (1997)]{malhotra} Malhotra, S. 1997, ApJ, 488, L101

\bibitem[Massa et al.(1983)]{massa} Massa, D., Savage, B. D., \& Fitzpatrick, E. L. 1983, ApJ, 266, 662

\bibitem[McGough et al.(2005)]{mcgough} McGough, C., Clayton, G. C., Gordon, K. D., \& Wolff, M. J. 2005, ApJ, 624, 118

\bibitem[Motta et al.(2002)]{motta} Motta, V. et al. 2002, ApJ, 574, 719

\bibitem[Mu\~noz et al.(2004)]{munoz} Mu\~noz, J. A., Falco, E. E., Kochanek, C. S., McLeod, B. A. \& Mediavilla, E. 2004, ApJ, 605, 614

\bibitem[Murphy \& Liske(2004)]{murphy} Murphy, M. T. \& Liske, J. 2004, MNRAS, 354, L31

\bibitem[Myers et al.(1999) ]{myers} Myers, S. T. \& Rusin, D. et al. 1999, AJ, 117, 6, 2565

\bibitem[Nadeau et al.(1991)]{nadeau} Nadeau, D., Yee, H. K. C., Forrest, W. J., Garnett, J. D., Ninkov, Z., \& Pipher, J. L. 1991, ApJ, 376, 430

\bibitem[Nair \& Garrett(1997)]{nair} Nair, S., \& Garrett, M.A. 1997, MNRAS,284, 58

\bibitem[Nandy et al.(1981)]{nandy81} Nandy, K., Morgan, D. H., Willis, A. J., Wilson, R., \& Gondhalekar, P. M. 1981, MNRAS, 196, 955

\bibitem[Patnaik et al.(1992)]{patnaik} Patnaik, A. R., Browne, I. W. A., Walsh, D., Chaffee, F. H., \& Foltz, C. B. 1992, MNRAS, 259, 1

\bibitem[Patnaik \& Narasimha(2001)]{patnar} Patnaik, A. R. \& Narasimha, D. 2001, MNRAS, 326, 1403


\bibitem[Pei et al.(1991)] {pei} Pei, Y. C, Fall, S. M. \& Bechtold, J. 1991, ApJ, 378, 6

\bibitem[Pei(1992)]{pei2} Pei, Y. C. 1992, ApJ, 395, 130

\bibitem[Perlmutter et al.(1997)]{perlmutter97} Perlmutter, S. et al. 1997, ApJ, 483,565

\bibitem[Perlmutter et al.(1999)]{perlmutter} Perlmutter, S. et al. 1999, ApJ, 517, 565

\bibitem[Pr\'evot et al.(1984)]{prevot84} Pr\'evot, M. L., Lequeux, J., Maurice, E., Pr\'evot, L., \& Rocca-Volmerange, B. 1984, A\&A, 132, 389

\bibitem[Remy et al.(1993)] {remy} Remy, M., Surdej, J., Smette, A., \& Claeskens, J.-F. 1993, A\&A, 278, L19

\bibitem[Riess et al.(1996a)]{riess} Riess, A. G., Press, W. H., \& Kirshner, R.P. 1996a, ApJ, 473, 88

\bibitem[Riess et al.(1996b)]{riess1996b} Riess, A. G., Press, W. H., \& Kirshner, R. P. 1996b, ApJ, 473, 588

\bibitem[Riess et al.(1998)]{riess1998} Riess, A. G. et al. 1998, AJ, 116, 1009

\bibitem[Rix et al.(1992)]{rix} Rix, H.-W., Schneider, D. P., \& Bahcall, J. N. 1992, AJ, 104, 959

\bibitem[Rusin et al.(2002a)]{rusin} Rusin, D., Norbury, M., Biggs, A. D., Marlow, D. R., Jackson, N. J., Browne, I. W. A., Wilkinson, P. N., \& Myers, S. T. 2002a, MNRAS, 330, 205

\bibitem[Rusin et al.(2002b)]{rusin_v1} Rusin, D. et al. 2002, astro-ph/0211229v1

\bibitem[Rusin et al.(2003)]{rusin2003} Rusin, D. et al. 2003, ApJ, 587, 143

\bibitem[Schechter et al.(1997)]{schechter} Schechter, P. L. et al. 1997, ApJ, 475, L85

\bibitem[Schneider et al.(1985)]{schneider85} Schneider, D. P., Lawrence, C. R., Schmidt, M., Gunn, J. E., Turner, E. L, Burke, B. F., \& Dhawan, V. 1985, ApJ, 294, 66

\bibitem[Schneider et al.(1986)]{schneider86} Schneider, D. P., Gunn., J. E., Turner, E. L., Lawrence, C. R., Hewitt, J. N., Schmidt, M., \& Burke, B. F. 1986, AJ, 91, 991

\bibitem[Schneider et al.(1988)]{schneider88} Schneider, D. P., Turner, E. L., Gunn, J. E., Hewitt, J. N., Schmidt, M., \& Lawrence, C. R. 1988, AJ, 95, 1619

\bibitem[Sullivan et al.(2003)]{sullivan} Sullivan, M. et al. 2003, MNRAS, 340, 1057

\bibitem[Surdej et al.(1987)]{surdej} Surdej, J. et al. 1987, Nature, 329, 695

\bibitem[Toft et al.(2000)]{toft} Toft, S., Hjorth, J., \& Burud, I. 2000, A\&A, 357, 115

\bibitem[Tonry(1998)]{tonry98} Tonry, J. L. 1998, AJ, 115, 1

\bibitem[Tonry \& Kochanek(1999)]{tonry} Tonry, J. L. \& Kochanek, C. S. 1999, AJ, 117, 2034

\bibitem[Udalski(2003)]{udalski} Udalski, A. 2003, ApJ, 590, 284

\bibitem[van Dokkum(2001)]{vandokkum} van Dokkum, P. G. 2001, PASP, 113, 1420

\bibitem[Vanden Berk et al.(2001)] {berk} Vanden Berk, D. E. et al. 2001, AJ, 122, 549

\bibitem[Wang(2005)]{wang} Wang, L. 2005, ApJ, 635, L33

\bibitem[Wang(2006)]{wang2006} Wang, X., Wang, L., Pain, R., Zhou, X., \& Li, Z. 2006, ApJ, in press (astro-ph/0603392)

\bibitem[Weynmann et al.(1980)]{weynman} Weynmann, R. J., Latham, D., Angel, J. R. P., Green, R. F., Liebert, J. W., Turnshek, D. A., Turnshek, D. E., \& Tyson, J. A. 1980, Nature, 285, L641

\bibitem[Wisotzki et al.(2004)]{wisotzki} Wisotzki, L., Becker, T., Christensen, L. et al. 2004, AN, 325, 135

\bibitem[Wucknitz et al.(2003)]{wucknitz} Wucknitz, O., Wisotzki, L., Lopez, S., \& Gregg, M. D. 2003, A\&A, 405, 445

\bibitem[Xanthopoulos et al.(1998)]{xan1} Xanthopoulos, E. et al. 1998, MNRAS 300, 649

\bibitem[Yee \& Ellingson (1994)]{yee} Yee, H. K. C. \& Ellingson, E. 1994, AJ, 107, 28

\bibitem[Yee(1988)] {yee88} Yee, H. K. C. 1988, AJ, 95, 1331

\bibitem[York et al.(2006)]{york} York, D. G. et al. 2006, MNRAS, 367, 945

\bibitem[\"Ostman et al.(2006)]{ostman} \"Ostman L. et al. 2006, A\&A, 450, 971


\end{thebibliography}
\end{document}